%% file: main_arxiv.tex
\documentclass{article}

\usepackage[preprint]{neurips_2025}

\usepackage[utf8]{inputenc} 
\usepackage[T1]{fontenc}    
\usepackage{hyperref}       
\usepackage{url}            
\usepackage{booktabs}       
\usepackage{amsfonts}       
\usepackage{nicefrac}       
\usepackage{microtype}      
\usepackage{xcolor}         
\usepackage{graphicx}
\usepackage{multirow}
\usepackage{amsmath}
\usepackage{colortbl}
\usepackage{cleveref}
\usepackage[ruled,vlined]{algorithm2e}

\usepackage[export]{adjustbox}
\usepackage{enumitem}

\newcommand\blfootnote[1]{%
  \begin{NoHyper}%
  \renewcommand\thefootnote{}\footnote{#1}%
  \addtocounter{footnote}{-1}%
  \end{NoHyper}%
}

\title{PixCell: A generative foundation model for digital histopathology images}
\newcommand{\sbu}{\textsuperscript{1}}
\newcommand{\anl}{\textsuperscript{2}}
\newcommand{\uoc}{\textsuperscript{3}}
\newcommand{\uou}{\textsuperscript{4}}

\author{%
    Srikar Yellapragada\textsuperscript{*}\sbu \And
    Alexandros Graikos\textsuperscript{*}\sbu \And
    Zilinghan Li\textsuperscript{$\dagger$}\anl \And
    Kostas Triaridis\textsuperscript{$\dagger$}\sbu \AND
    Varun Belagali\sbu \And
    Tarak Nath Nandi\anl\textsuperscript{,}\uoc \And
    Karen Bai\sbu \And 
    Beatrice S. Knudsen\uou \And
    Tahsin Kurc\sbu \AND 
    Rajarsi R. Gupta\sbu \And
    Prateek Prasanna\sbu \And
    Ravi K Madduri\anl\textsuperscript{,}\uoc \And
    Joel Saltz\sbu \And
    Dimitris Samaras\sbu 
}
\date{%
    \textbf{\sbu Stony Brook University} \AffilAnd
    \hfill \textbf{\anl Argonne National Laboratory} \AffilAnd
    \hfill \textbf{\uoc  The University of Chicago} \AffilAnd
    \hfill \textbf{\uou  University of Utah}\hfill%
}

\begin{document}

\maketitle

\begin{center}
    \vspace{-0.0cm}
    \href{https://histodiffusion.github.io/docs/projects/pixcell/}{
        \adjustbox{valign=c}{%
          \includegraphics[height=1.2\baselineskip]{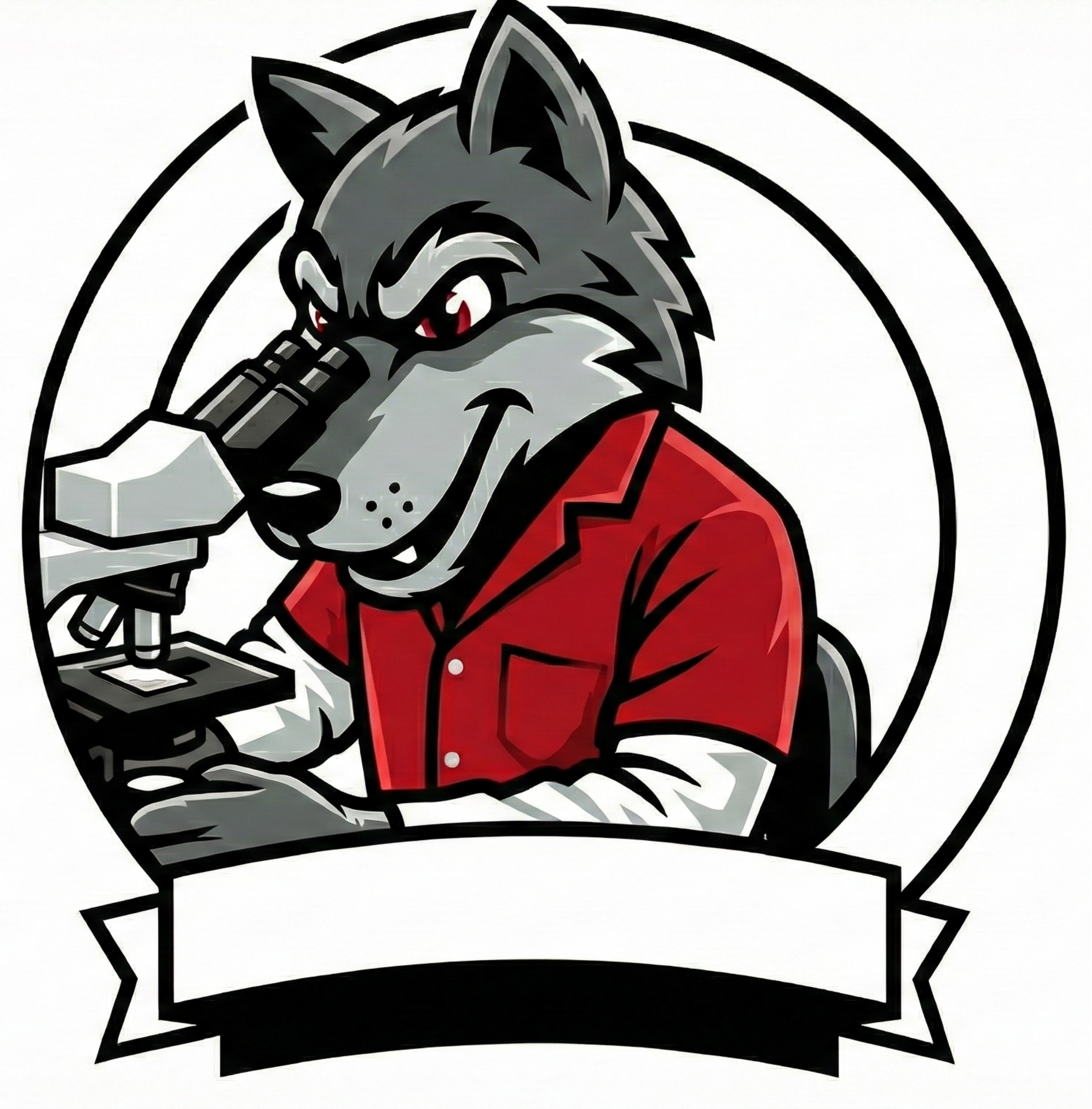}%
        }
        Project Page
    }
    \hskip 0.2in
    \href{https://huggingface.co/StonyBrook-CVLab/PixCell-256}{
        \adjustbox{valign=c}{%
          \includegraphics[height=1.2\baselineskip]{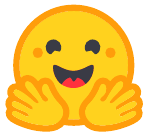}%
        }
        PixCell-256
    }
    \hskip 0.2in
    \href{https://huggingface.co/StonyBrook-CVLab/PixCell-1024}{
        \adjustbox{valign=c}{%
          \includegraphics[height=1.2\baselineskip]{figures/huggingface_logo-noborder.pdf}%
        }
        PixCell-1024
    }
    \hskip 0.2in
    \href{https://huggingface.co/datasets/StonyBrook-CVLab/Synthetic-TCGA-10M}{
        \adjustbox{valign=c}{%
          \includegraphics[height=1.2\baselineskip]{figures/huggingface_logo-noborder.pdf}%
        }
        Synthetic TCGA-10M
    }
\end{center}

\begin{abstract}
The digitization of histology slides has revolutionized pathology, providing massive datasets for cancer diagnosis and research. Self-supervised and vision-language models have been shown to effectively mine large pathology datasets to learn discriminative representations. On the other hand, there are unique problems in pathology, such as annotated data scarcity, privacy regulations in data sharing, and inherently generative tasks like virtual staining. Generative models, capable of synthesizing realistic and diverse images, present a compelling solution to address these problems through image synthesis. 
We introduce \textbf{PixCell, the first generative foundation model for histopathology images}. PixCell is a diffusion model trained on PanCan-30M, a large, diverse dataset derived from 69,184 H\&E-stained whole slide images of various cancer types. We employ a progressive training strategy and a self-supervision-based conditioning that allows us to scale up training without any human-annotated data.
PixCell generates diverse and high-quality H\&E-stained images of multiple cancer types, conditioned on embeddings extracted from real slides. By conditioning on real slides, the synthetic images capture the properties of the real data and can be used as data augmentation for small-scale datasets to boost classification performance.
We prove the foundational versatility of PixCell by applying it to two generative downstream tasks: \textit{privacy-preserving synthetic data generation} and \textit{virtual IHC staining}.
PixCell's high-fidelity conditional generation enables institutions to use their private data to synthesize highly realistic, site-specific surrogate images that can be shared in place of raw patient data. Other sites can then use these synthetic images to augment their local datasets with examples that would otherwise be unavailable, including rare disease presentations and cohorts with distinct specimen preparation or staining protocols. Furthermore, using datasets of roughly paired H\&E–IHC tiles, we learn to translate PixCell's conditioning from H\&E to multiple IHC stains, allowing the generation of IHC images from H\&E inputs. The virtually-stained images achieve state-of-the-art perceptual scores and strong agreement with both automated scoring and expert pathologists. Our trained models are publicly released to accelerate research in computational pathology.

\blfootnote{\textsuperscript{*} Co-first authors  \textsuperscript{$\dagger$} Co-second authors. Correspondence at \{myellapragad, agraikos, samaras\}@cs.stonybrook.edu}

\end{abstract}


\input{Sections/introduction}

\input{Sections/results}

\input{Sections/method}
\input{Sections/experimental_setup}
\input{Sections/related_work}
\input{Sections/conclusion}


\bibliographystyle{abbrv}
\bibliography{references}


\input{Sections/appendix}


\end{document}

%% file: Sections/introduction.tex
\section{Introduction}
Computational pathology enables the extraction and quantitative analysis of rich information available in digitized tissue samples. It has brought significant advances in studies of disease mechanisms, disease classification (such as tumor subtyping) \cite{bracs, fu2020pan, saltz2018spatial}, segmentation and characterization of important spatial features in diseased tissue \cite{graham2019hover, sirinukunwattana2017gland}, and predictive analytics for disease diagnosis, staging, and prognosis. Most modern computational pathology approaches employ Artificial Intelligence (AI) methods. Pathology AI models have demonstrated superior performance compared to other analysis techniques; however, developing robust models is often hindered by data-related challenges. Generating pixel- and patch-level annotations for training data is typically laborious, requires the involvement of expert pathologists, and is therefore time-consuming and expensive \cite{Hou_2019_CVPR}.   

\begin{figure}[p]
    \centering
    \includegraphics[width=\linewidth]{figures/overview.pdf}
    \caption{\textbf{PixCell overview.} \textbf{a.} Our training data comprises a large collection of 69,184 WSIs spanning 28 tissue types. \textbf{b.} We progressively increase image resolution during training: We start by training the model on 256x256 images, conditioning the generation process on the image embedding extracted from a pretrained pathology foundation model (Stage 1). We continue training the same model with 512x512 (Stage 2) and 1024x1024 images (Stage 3), using embeddings from all 256x256 tiles contained in the larger image. \textbf{c.} PixCell generates images that preserve key features of the reference tiles and are perceived as highly similar by pre-trained pathology image encoders. Using synthetic images for data augmentation with these images improves the performance of downstream classifiers. Using an inference-time algorithm, we can scale the generated images to 4096x4096 pixels. \textbf{d.} Synthetic images can serve as a drop-in replacement for real data in the training of self-supervised foundation models. This enables privacy-preserving data sharing between institutions. \textbf{e.} Using PixCell with small datasets of paired H\&E and IHC images enables virtual staining. Our virtual staining pipeline leads to higher accuracy in the diagnostic labels predicted by both automated and human evaluators. Figure best viewed in $3\times$ magnification.}
    \label{fig:overview}
\end{figure}

To address these data challenges, the prevailing approach has been to train powerful self-supervised learning (SSL) models from vast amounts of unlabeled data; these models can be adapted to perform downstream tasks \cite{uni, virchow2, hoptimus1, hibou, plip, quilt, conch, musk}. Models trained via an SSL method learn to extract salient and discriminative embeddings from images; these embeddings can be utilized, with or without fine-tuning, in particular downstream tasks. Studies have shown that such models encode useful diagnostic information~\cite{uni, hoptimus1, conch}.

Although SSL-trained discriminative embeddings solve several computational pathology tasks, their use does not address all data challenges. Firstly, using a pre-trained self-supervised model to solve a downstream task may involve training a prediction layer; the training process may require less, but still a non-trivial amount of high-quality annotated data. Secondly, SSL training benefits from curating diverse training datasets. Strict patient privacy rules and concerns over data ownership frequently limit data sharing. As a result, in most cases, institutions or consortia end up training their own SSL model primarily with private data supplemented by a few commonly used public datasets such as the TCGA whole slide imaging dataset \cite{weinstein2013cancer}. Data sharing challenges limit model performance and can potentially create bias due to the specific demographics of training datasets \cite{vaidya2024demographic, kheiri2025investigation}. 
Finally, discriminative models, trained to discern between different images, are not suitable for tasks that require synthesizing a completely new image from a reference image, such as virtual staining (stain translation) \cite{boyd2022region, pati2024accelerating, li2023adaptive}.

Generative models, capable of controlled synthesis of realistic tissue images, present a compelling solution to rare disease presentations, annotated data scarcity, and the need for privacy-preserving data sharing, as well as to carry out inherently generative tasks, like virtual staining. We introduce \textbf{PixCell}, the first generative foundation model for histopathology images. We employ the Diffusion Transformer (DiT) architecture \cite{peebles2023scalable}, and train PixCell on \textbf{PanCan-30M} -- an extensive dataset of 30.8 million $1024\times1024$ px image patches derived from 69,184 H\&E-stained whole slide images (WSIs), covering a comprehensive range of cancer types and benign tissues. To effectively scale up training we employ: (a) progressive training, starting from $256\times256$ tiles (PixCell-256) and gradually increasing the generated image size to $1024\times1024$ (PixCell-1024) and (b) per-patch guidance, by conditioning PixCell on feature embeddings from a pre-trained self-supervised model (here we chose UNI-2h \cite{uni}) which provides rich image semantics to guide the generation process.

We show that PixCell generates synthetic images of unprecedented quality, both in terms of their \textit{realism} and the diversity of the \textit{biologically plausible features} seen in real tissue. 
We utilize PixCell's synthetic images for data augmentation, which boosts the performance of downstream diagnostic models by an average of $3\%$ across multiple datasets and architectures. We prove the foundational versatility of PixCell by applying it on two generative downstream tasks: privacy-preserving synthetic data generation and virtual IHC staining.

We demonstrate that our synthetic data can serve as a drop-in replacement for real data in the training of foundation models. An SSL encoder trained exclusively on synthetic TCGA images achieves performance comparable to that of an encoder trained on real TCGA images. We also provide a proof-of-concept for privacy-preserving data sharing: training a model on real data from one data source (TCGA) and synthetic data from another data source (HistAI \cite{histai}) yields a superior model that outperforms the baseline trained on data from a single institution. We publicly release the synthetic TCGA dataset to facilitate further research on synthetic histology data.

Furthermore, PixCell demonstrates broad generalization capabilities, successfully synthesizing images with staining techniques unseen during training. We investigate PixCell's applicability to virtual staining by learning to translate H\&E images to various IHC stains using small datasets of approximately paired H\&E and IHC tiles. PixCell improves the state-of-the-art in inferring IHC stains from an H\&E-stained image. The synthesis of images that are out of the training domain of PixCell is further proof of its versatility as a foundation model that can be efficiently adapted to specialized generative tasks using limited additional data.

%% file: Sections/results.tex
\begin{figure}[t]
    \centering
    \includegraphics[width=1\linewidth]{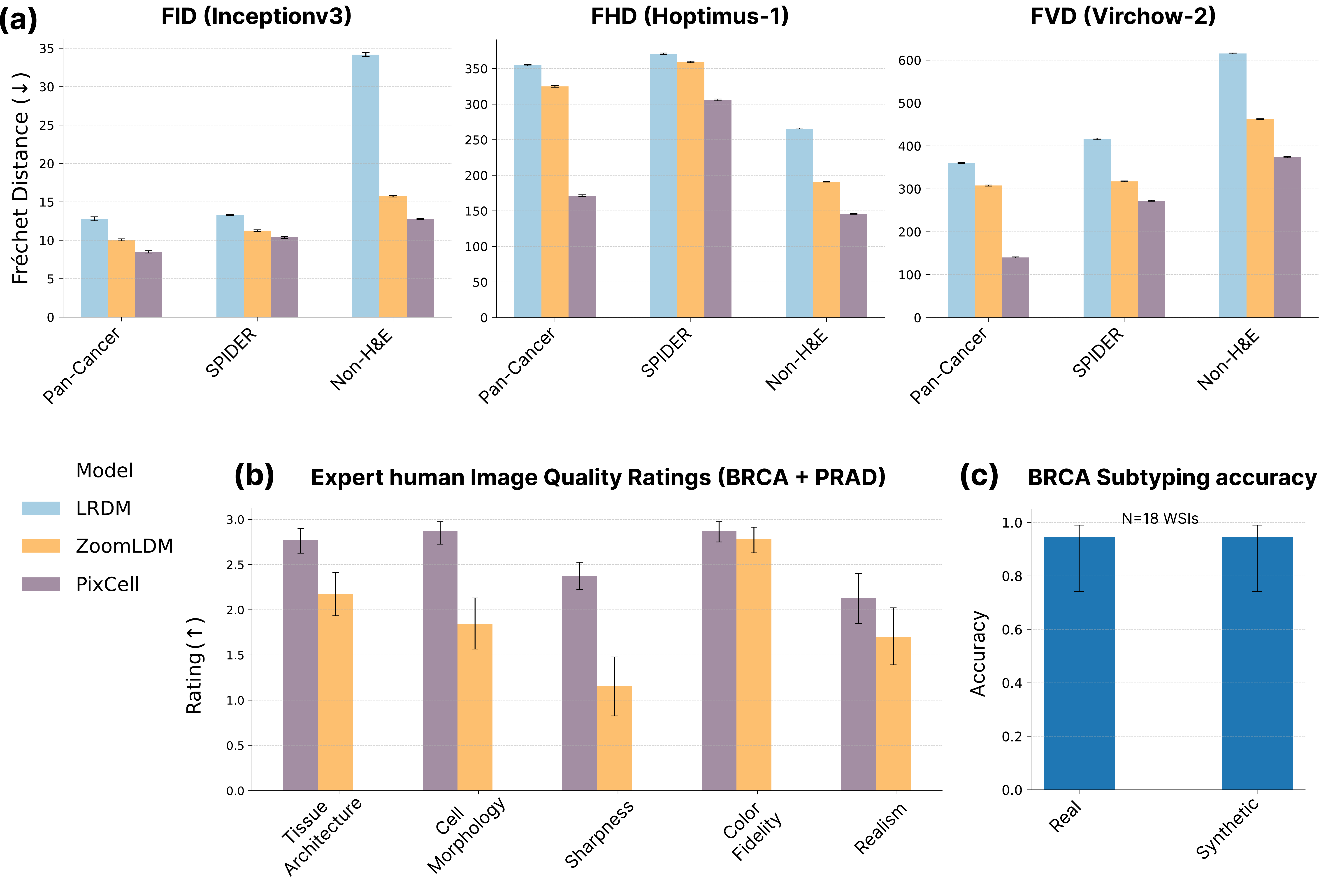}
    \caption{\textbf{a.} PixCell consistently achieves lower (better) Fréchet Distance scores across multiple datasets and pathology-specific encoders. \textbf{b.} Two expert pathologists rated PixCell's synthetic images as having higher fidelity across all five qualitative criteria. \textbf{c.} Pathologist prediction accuracy for breast cancer subtyping (lobular vs ductal) on synthetic images (N=18 WSIs) is nearly identical to that on real images. Error bars denote 95\% Confidence Interval (CI). }
    \label{fig:pixcell_image_quality}
\end{figure}

\section{Results}

\subsection{PixCell generates high-quality realistic pathology images}
\label{sec:results_gen_256}

We first evaluate whether PixCell can synthesize high quality histopathology images, both in terms of their realism and the diversity of the biologically plausible features seen in real tissue.  We assess generation quality by benchmarking PixCell against prior diffusion-based methods (LRDM \cite{cvpr24_ldm} and ZoomLDM \cite{zoomldm}) using Fréchet Distance (FD) \cite{heusel2017gans} with multiple feature extractors - Inceptionv3 \cite{szegedy2016rethinking}, Hoptimus-1 \cite{hoptimus1} and Virchow-2 \cite{virchow2}. The use of multiple feature extractors allows us to evaluate realism both in a generic vision sense and in a pathology-aware sense. Our evaluation is performed on three datasets that test distinct aspects of generalization: \textbf{PanCan-Test}, a held-out set from our training distribution; \textbf{SPIDER}, which contains images from an unseen imaging center; and \textbf{Non-H\&E}, which contains images prepared with alternative staining protocols.

PixCell consistently achieves the lowest (best) Fréchet Distance across all three datasets (Fig. \ref{fig:pixcell_image_quality} a). The advantage is more pronounced for pathology-specific feature spaces \cite{hoptimus1, virchow2}; for instance, on the Pan Cancer test set, PixCell-256 achieves an FHD (Hoptimus-1) score nearly 50\% lower than the next best model. This demonstrates a superior alignment with pathology-specific features. Furthermore, the strong performance on the out-of-distribution SPIDER and Non-H\&E datasets confirm that PixCell generalizes effectively to unseen data sources and staining variations, despite being trained exclusively on H\&E slides. Quantitative performance metrics for our high-resolution PixCell-1024 model are detailed in Appendix Section \ref{sec:high_res_supp}.

Going beyond automated metrics, we evaluate how accurately expert pathologists can interpret large images generated by PixCell. As shown in Fig. \ref{fig:pixcell_image_quality} b, across both BRCA and PRAD, PixCell consistently received higher image-quality scores from expert pathologists across all five criteria. Furthermore, the pathologist’s subtype predictions on synthetic BRCA regions matched real-image decisions with a 94.4\% accuracy (Fig. \ref{fig:pixcell_image_quality} c). The decisions were fully concordant in every case except one (17 out of 18 WSIs), where the synthetic regions were labeled ``Unsure." This demonstrates that PixCell-generated images preserve sufficient and accurate morphological detail for clinical interpretation.

\begin{figure}[t]
    \centering
    \includegraphics[width=1\linewidth]{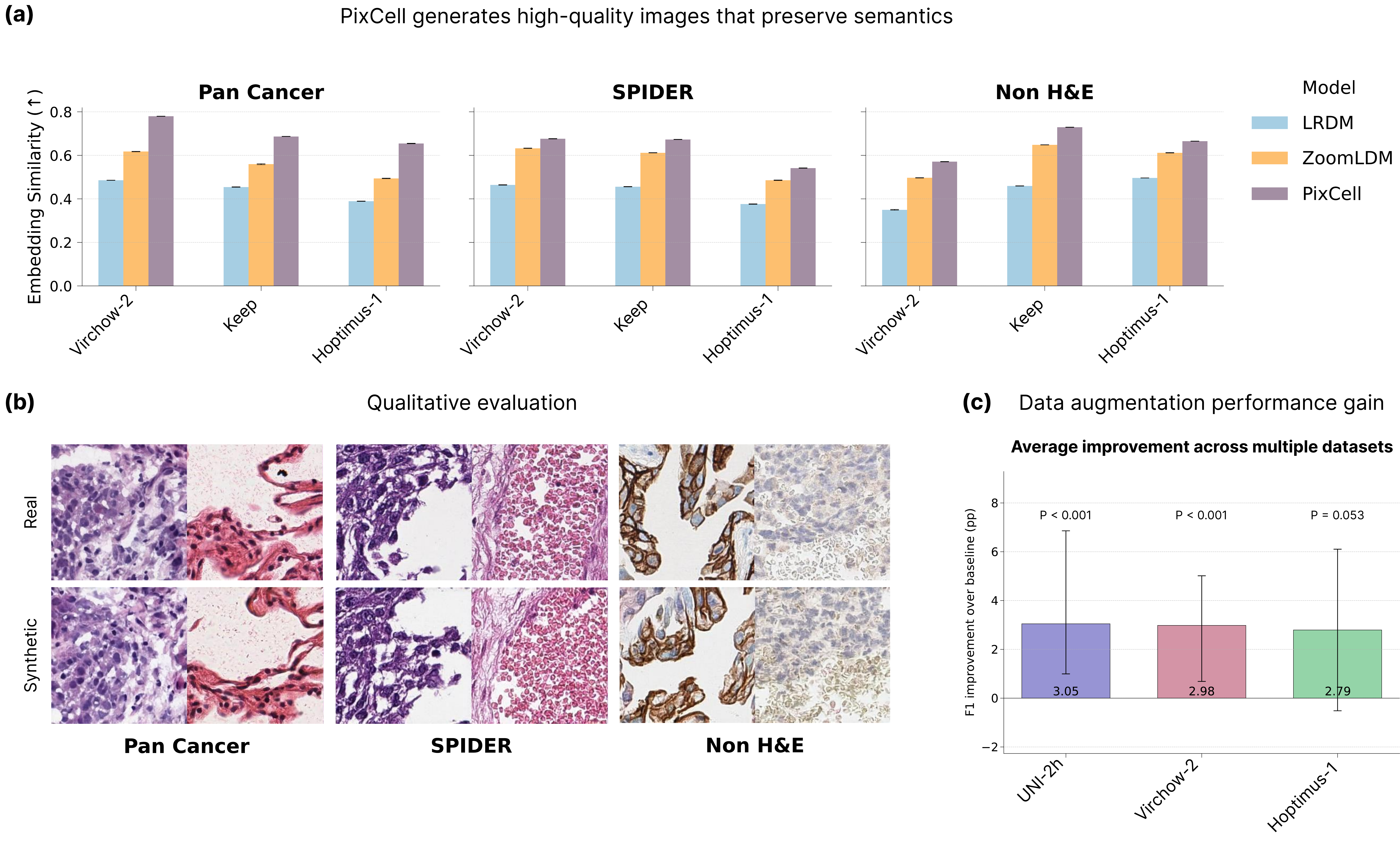}
    \caption{(a) The embeddings of PixCell's synthetic images are very similar to the embeddings of real images across multiple encoders which leads to (b) synthetic images preserving tissue semantics  (c) Augmenting training data with synthetic images consistently improves the F1 scores of the three state-of-the-art downstream classifiers. Error bars represent $95\%$ CI.}
    \label{fig:augmentations}
\end{figure}

\subsection{Synthetic images preserve semantics and improve downstream model performance}
\label{sec:results_aug}

Previous models (LRDM and ZoomLDM) also use image embeddings from a reference image to generate new images. Beyond distribution distance, a model's utility depends on its ability to faithfully preserve the semantic content of the reference image. To assess this ability, we measure the cosine similarity between the real and synthetic patches in the embedding space of three large pathology foundation encoders, Hoptimus1 \cite{hoptimus1}, KEEP \cite{keep} and Virchow-2 \cite{virchow2} across multiple datasets. In all cases, PixCell produced the highest embedding similarity to real data (Fig \ref{fig:augmentations} a), reaching similarity scores above 0.7.

The high embedding similarity confirms that PixCell generates novel variations of an image while successfully preserving key morphological features (see Fig  \ref{fig:augmentations} b for a qualitative evaluation). This allows us to expand training datasets with diverse, yet semantically consistent examples. To evaluate this capability, we augment the training data of four small, patch-level classification datasets (BACH \cite{bach}, BRACS \cite{bracs}, Break-his \cite{breakhis} and mHist \cite{mhist}) with synthetic samples. Due to limited number of examples in these datasets, pretrained image encoders tend to overfit (see Appendix \ref{sec:supp_data_aug}). We generate synthetic data for each of these datasets and then measure the change in performance for three classifiers built on top of state-of-the-art encoders (UNI-2 \cite{uni}, Virchow-2 \cite{virchow2}, and Hoptimus-1 \cite{hoptimus1}) against baselines trained only on real data.

We determine the optimal augmentation strategy (Vanilla vs. Test-time) for each encoder-dataset pair via a grid search on the validation set (details in Sec \ref{sec:method_data_aug} and \ref{sec:setup_data_aug}). As shown in Fig. \ref{fig:augmentations} c, when averaged across datasets, our augmentation strategy yields a consistent F1 score improvement of approximately $3 \%$. This demonstrates that PixCell-generated images display diverse pathologic entities, while also retaining the important semantic features that are used in properly classifying them into distinct categories.

\subsection{PixCell enables privacy-preserving data sharing}
\label{sec:syn_ssl}

While data augmentation can improve the performance of existing encoders, training powerful foundation models still faces a fundamental challenge: institutions are often unable or unwilling to share data across sites because of privacy rules, IRB requirements, and proprietary concerns. PixCell's ability to generate synthetic images with feature  distributions and semantic context that are highly aligned with real data allows us to generate synthetic variants of entire pathology datasets and use them in place of real data. Although this does not guarantee complete data privacy, as generative models have been prone to leaking training samples \cite{carlini2023extracting}, it is the first step towards enabling synthetic data sharing for model training. While our generation process is conditioned on embeddings from real data, the output consists entirely of synthetic samples, which decouples the shared data from the original raw patient scans.

First, to validate the quality of our synthetic data, we test if it can serve as a direct substitute for real data. We train two separate DINOv2 \cite{dinov2} encoders: one exclusively on 10 million real patches from TCGA \cite{weinstein2013cancer} (TCGA Real), another on a synthetic copy generated by PixCell (TCGA Synthetic). When evaluated on 9 downstream tasks, the TCGA Synthetic model achieves an average balanced accuracy of $76.64 \%$, which is close to the $77.89 \%$ achieved by the TCGA Real model (Table \ref{tab:syn_ssl}). This demonstrates that PixCell-generated images can act as high-fidelity drop-in replacement for real images during SSL pre-training.

Next, we demonstrate a proof-of-concept for data sharing between two institutions that cannot directly exchange real slides. We simulate a scenario where Institute 1, which has access to the TCGA data, collaborates with Institute 2, which has its own internal dataset, HistAI \cite{histai}. Instead of sharing the real HistAI data, we generate a synthetic, 10 million variant of HistAI. Institute 2 shares this synthetic data and we can train a new model on the combined dataset of Institute 1's real TCGA and synthetic HistAI patches (TCGA Real + HistAI syn).

We simulate a scenario where Institute 1, which has access to the TCGA data, collaborates with Institute 2, which has its own internal dataset, HistAI \cite{histai}. Instead of sharing the real HistAI data, we generate a synthetic, 10 million variant of HistAI. Institute 2 shares this synthetic data and we can train a new model on the combined dataset of Institute 1's real TCGA and synthetic HistAI patches (TCGA Real + HistAI syn). This model achieves an average accuracy of $79.11 \%$, outperforming the baseline model trained only on the real TCGA data. This finding establishes a viable method for institutions to pool their datasets by sharing synthetic copies, without transferring the original, sensitive WSI data. We investigate the scaling properties of synthetic data in Appendix Section \ref{sec:syn_ssl_scaling}. We release the Synthetic TCGA-10M dataset and invite the community to experiment with our synthetic data.

\begin{table}[ht]
    \centering
    \caption{k-NN performance of SSL encoders (DINOv2 ViT-B) trained on real and synthetic patches from TCGA, and a combination of TCGA real + HistAI synthetic patches. We use the Thunder framework~\cite{thunder} to evaluate on patch classification datasets and report balanced accuracy.}
    \label{tab:syn_ssl}

\input{tables/knn_tcga_10m}

\end{table}

\subsection{PixCell performs diagnostically accurate virtual staining}
\label{sec:results_virtual_staining}

PixCell is trained to synthesize H\&E-stained image variations using embeddings extracted from a reference H\&E tile. We find that our large-scale training allows the model to generalize to other, unseen stains, synthesizing plausible variations, as long as the conditioning encoder accurately embeds the reference image. We showcased this in Figure~\ref{fig:pixcell_image_quality}, where we showed that our model significantly outperforms baselines on non-H\&E stained datasets, and visually demonstrate it in Figure~\ref{fig:pixcell_ihc_variations}.

PixCell's ability to generalize to unseen staining techniques motivates us to utilize it as a strong generative prior for virtual staining tasks; given an H\&E-stained tile, the goal is to synthesize plausible images of the same tissue region, as if it was stained with a different method (e.g. IHC stains). This is an inherently generative task, requiring the synthesis of a new tissue image from a reference tile. We speculate that large-scale generative pre-training is critical in accurately performing virtual staining. To validate this hypothesis, we employ two datasets of H\&E and IHC images, MIST \cite{li2023adaptive} and HER2Match \cite{klockner2025gans}. The MIST slides are extracted from nearby slices of the same tissue block, resulting in loosely-paired image data, whereas HER2Match utilizes a re-staining technique to provide pixel-accurate H\&E and IHC pairs.

Evaluating the virtual staining task of H\&E to IHC is inherently challenging, as the evaluation metrics must appropriately capture the clinical properties of the generated images. Pixel-level metrics, like PSNR and SSIM, over-rely on local alignment between the generated and real images, which is not directly correlated to diagnostic accuracy. Instead of pixel-level, we evaluate our virtual staining using perceptual metrics (Fréchet Distance) that capture differences at the high-level feature level, and an automatic diagnostic evaluation scheme that utilizes the virtually-stained images to make diagnostic predictions. For feature extractors, we utilize both the standard (Inceptionv3 - FID) and pathology-specific (Hoptimus-1 - FHD, Virchow2 - FVD) image encoders, whereas for making diagnostic predictions, we use DeepLIIF \cite{deepliif}.

\begin{figure}[t]
    \centering
    \includegraphics[width=1\linewidth]{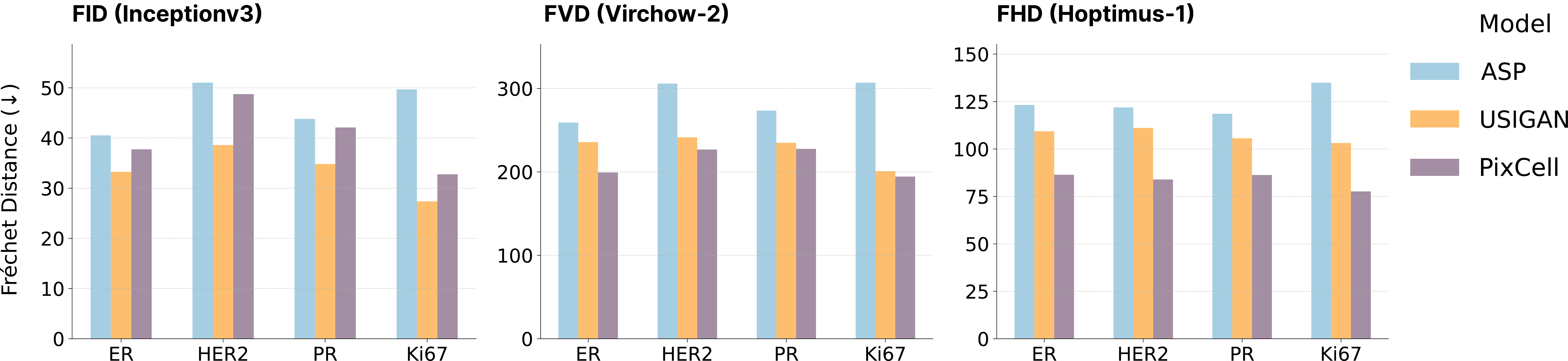}
    \caption{Fréchet distance between the generated and real IHC tiles using different encoders. For pathology-specific encoders, we find that PixCell achieves the best scores.}
    \label{fig:mist_frechet}
\end{figure}

\begin{figure}[t]
    \centering
    \includegraphics[width=0.9\linewidth]{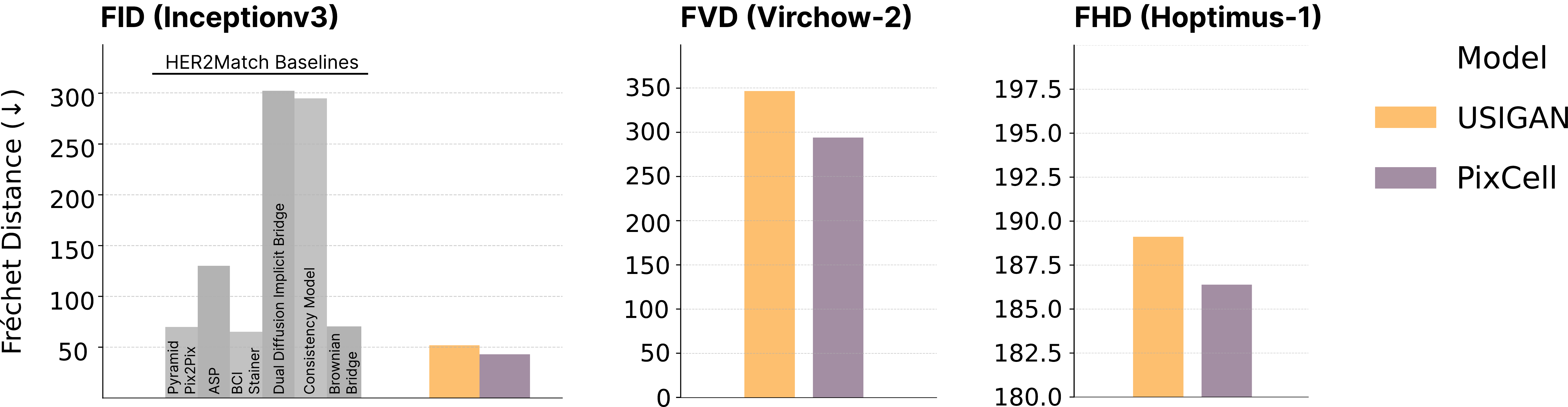}
    \caption{Fréchet distance between the generated and real IHC tiles using different encoders. Previous methods benchmarked by Klockner et al. \cite{klockner2025gans} perform worse than the USIGAN baseline. PixCell outperforms all previous methods when measuring distance with pathology-specific encoders.}
    \label{fig:her2match_frechet}
\end{figure}

On the MIST dataset, we compare our model to the CycleGAN \cite{zhu2017unpaired} baselines ASP \cite{li2023adaptive} and USIGAN \cite{peng2025usigan}. The comparison is separate for each stain (HER2, ER, PR,  Ki67). In Figure~\ref{fig:mist_frechet}, we show that PixCell outperforms both baselines when measuring perceptual distance with pathology-specific image encoders. For the HER2Match dataset, we first report the FID evaluations performed by Klockner et al. \cite{klockner2025gans} (Figure~\ref{fig:her2match_frechet} left). There, we show that PixCell outperforms all baselines reported. To compare with a state-of-the-art virtual staining method, we retrain the USIGAN baseline, which also beats all previously-reported FID scores. In Figure~\ref{fig:her2match_frechet} we show that PixCell again achieves the best pathology-specific perceptual scores.

For the diagnostic evaluation of virtually-stained IHC images, we assess the expression of the ER, PR, and Ki-67 stains on the MIST dataset. We utilize standardized protocols and guidelines \cite{path1, path2, path3, path4} that label IHC images based on the percentage of stained tumor cell nuclei, as also proposed by Klockner et al. \cite{klockner2025h}. We automatically count the number of stained and total cell nuclei in the real and generated IHC images using the DeepLIIF model\cite{deepliif}, and compare the labels assigned to the generated and real images for all images in the MIST test set. To verify the validity of the automated evaluation, we also asked an expert pathologist to hand-label a subset of real and synthetic IHC images and measured the alignment between the two scoring schemes.

\begin{figure}[t]
    \centering
    \includegraphics[width=1.0\linewidth]{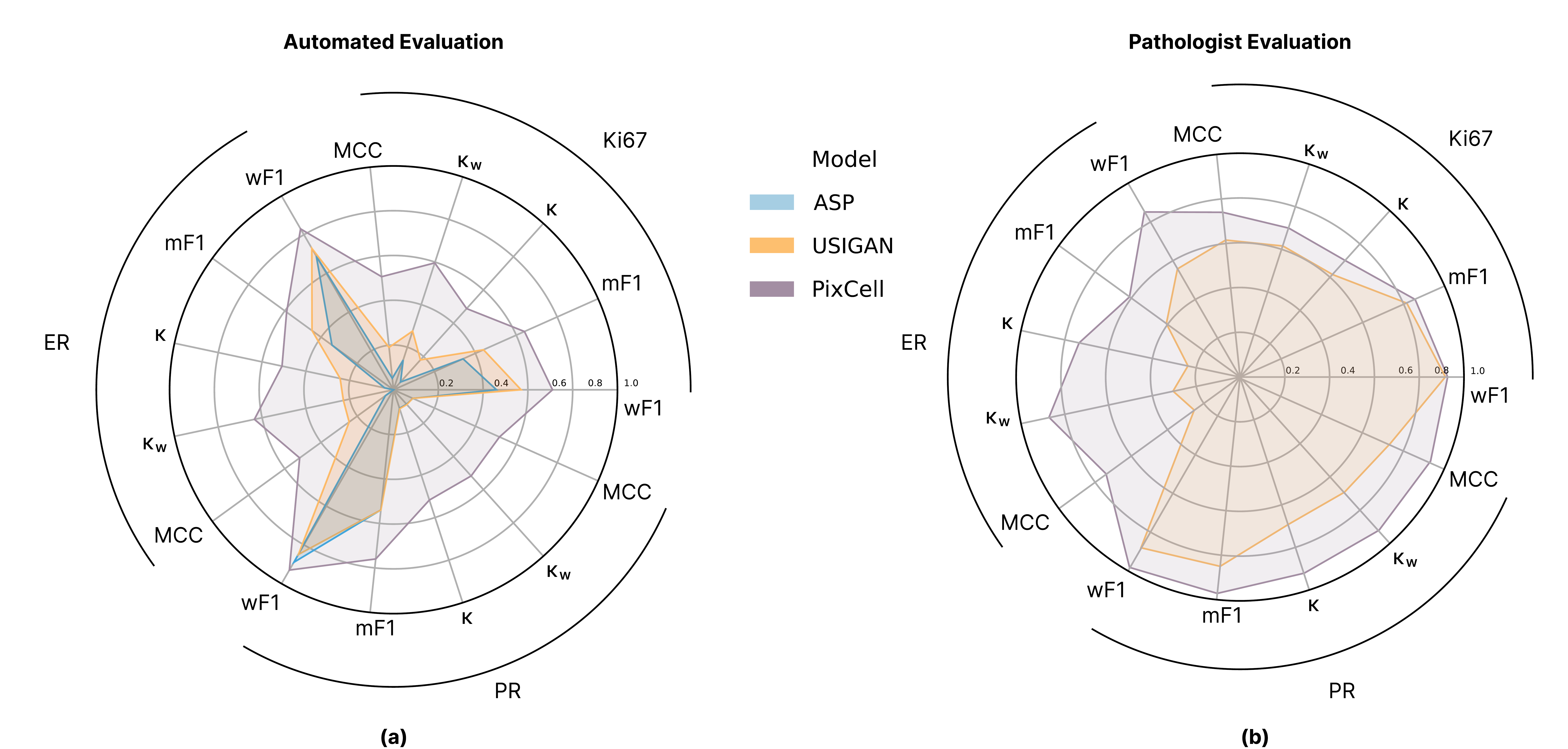}
    \caption{PixCell outperforms prior methods in both automated and pathologist scoring of virtually-stained IHC images.}
    \label{fig:scoring}
\end{figure}

\begin{figure}[t]
    \centering
    \includegraphics[width=1\linewidth]{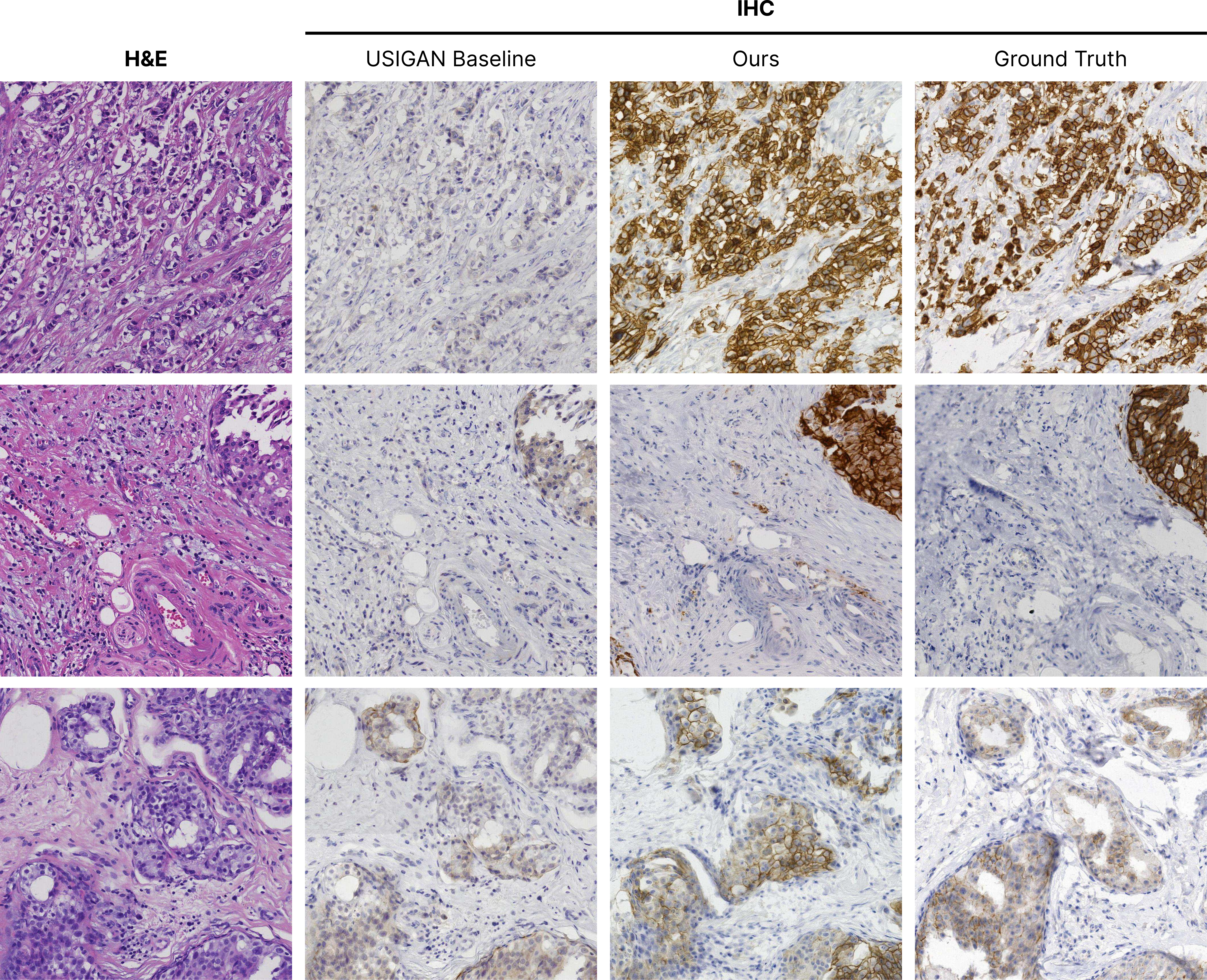}
    \caption{Results for H\&E \textrightarrow IHC stain translation on MIST using PixCell and a CycleGAN baseline (USIGAN). The baseline fails to translate critical features between the H\&E and IHC images.}
    \label{fig:stain_translation_baseline}
\end{figure}

The results of the automated evaluation for PixCell, ASP, and USIGAN are shown in Figure~\ref{fig:scoring} (a). We report the weighted F1 (wF1) and macro F1 (mF1), Cohen's kappa ($\kappa$) and weighted kappa ($\kappa_w$), and Matthews correlation coefficient (MCC). We find that PixCell substantially outperforms both baselines, showing that virtually-stained images with PixCell can be used to make diagnostic predictions. For the pathologist evaluation (Figure~\ref{fig:scoring} (b)), we only presented examples from PixCell and USIGAN and again find that our method is significantly better at synthesizing images that accurately translate the cell features from H\&E to the target IHC stain. In a qualitative comparison in Figure~\ref{fig:stain_translation_baseline}, we observe that PixCell-stained images accurately translate the H\&E images into the target IHC stain, whereas previous methods failed at capturing diagnostically-critical features.

%% file: tables/knn_tcga_10m.tex
\resizebox{\textwidth}{!}{
\setlength{\tabcolsep}{4pt}
\begin{tabular}{c|c|c|c|cccc|c|c|c}
\toprule
\multirow{2}{*}{Model} & \multirow{2}{*} {\begin{tabular}[c]{@{}c@{}}bracs \\ ~\cite{bracs}\end{tabular}}  & \multirow{2}{*} {\begin{tabular}[c]{@{}c@{}}ccrcc \\ ~\cite{brummer2023computational}\end{tabular}} & \multirow{2}{*}{\begin{tabular}[c]{@{}c@{}}crc \\ ~\cite{nct-crc}\end{tabular}} & \multicolumn{4}{c|}{SPIDER ~\cite{nechaev2025spider}} & \multirow{2}{*}{\begin{tabular}[c]{@{}c@{}}tcga crc \\ msi ~\cite{kather2020histological}\end{tabular}} & \multirow{2}{*}{\begin{tabular}[c]{@{}c@{}}tcga \\ tils ~\cite{komura2020histology}\end{tabular}} & \multirow{2}{*}{Avg} \\
                       &                        &                        &                      & Breast & Colorec & Skin  & Thorax &                                                                          &                            &                      \\ \midrule
TCGA Real              & 53.03                  & 80.48                  & 91.54                & 77.19  & 82.43   & 85.13 & 86.81  & 59.79                                                                    & 84.59                      & 77.89                \\
TCGA Synthetic              & 50.98                  & 75.03                  & 92.69                & 75.52  & 81.65   & 84.26 & 84.32  & 61.18                                                                    & 84.18                      & 76.64                \\
TCGA Real + HistAI Syn & 55.03                  & 77.99                  & 93.16                & 76.33  & 85.64   & 84.15 & 91.11  & 63.18                                                                    & 85.37                      & \textbf{79.11}                \\ \bottomrule
\end{tabular}
}

%% file: Sections/method.tex
\section{Method and Data}

\subsection{Training dataset: PanCan-30M}
Our training cohort comprises 69,184 Whole Slide Images (WSIs), aggregated to capture a wide variety of tissue morphologies and cancer subtypes. Inspired by Phikon-v2~\cite{phiconv2}, we source the WSIs from major public repositories including The Cancer Genome Atlas (TCGA) \cite{weinstein2013cancer}, the Clinical Proteomic Tumor Analysis Consortium (CPTAC) \cite{edwards2015cptac}, non-cancer tissue from the Genotype-Tissue Expression (GTEx) project \cite{lonsdale2013genotype}, and other public sources \cite{chowdhury2023proteogenomic, national2011reduced}, alongside an internal cohort (SBU). The pan-cancer nature of this WSI collection, covering a wide variety of organs, provides a diverse cohort for training a generative foundation model. In Figure~\ref{fig:overview} (a) we provide a breakdown of the WSIs per organ type used in training.

From these WSIs, we extract 30 million (30,819,977) patches, creating \textbf{PanCan-30M}. Each patch is $1024 \times 1024$ pixels in size, and is obtained from slide regions corresponding to $20 \times$ objective magnification (0.5 microns per pixel). We use the code from DS-MIL \cite{li2021dual} for both patch extraction and tissue thresholding. The slides included in the dataset are stained with Hematoxylin and Eosin (H\&E), and we do not use stain normalization. See Table~\ref{tab:dataset} for an exact breakdown of the data sources for the extracted patches. From this data, we create a separate hold-out test set of 2.3 million image patches, which we call \textbf{PanCan-Test}. In Appendix Figure~\ref{fig:data_patches} and Table~\ref{tab:dataset_detailed} we provide an exhaustive breakdown of the dataset per organ and data source.

\begin{table}[ht]
    \centering
    \caption{Details of the data sources used in the PanCan-30M dataset. We extract the $1024\times1024$ patches from whole-slide images across various organs from each data source.}
    \input{tables/data_sources}
    \label{tab:dataset}
\end{table}

\subsection{PixCell diffusion model}
\label{sec:training}

\begin{figure}[ht]
    \centering
    \includegraphics[width=0.8\linewidth]{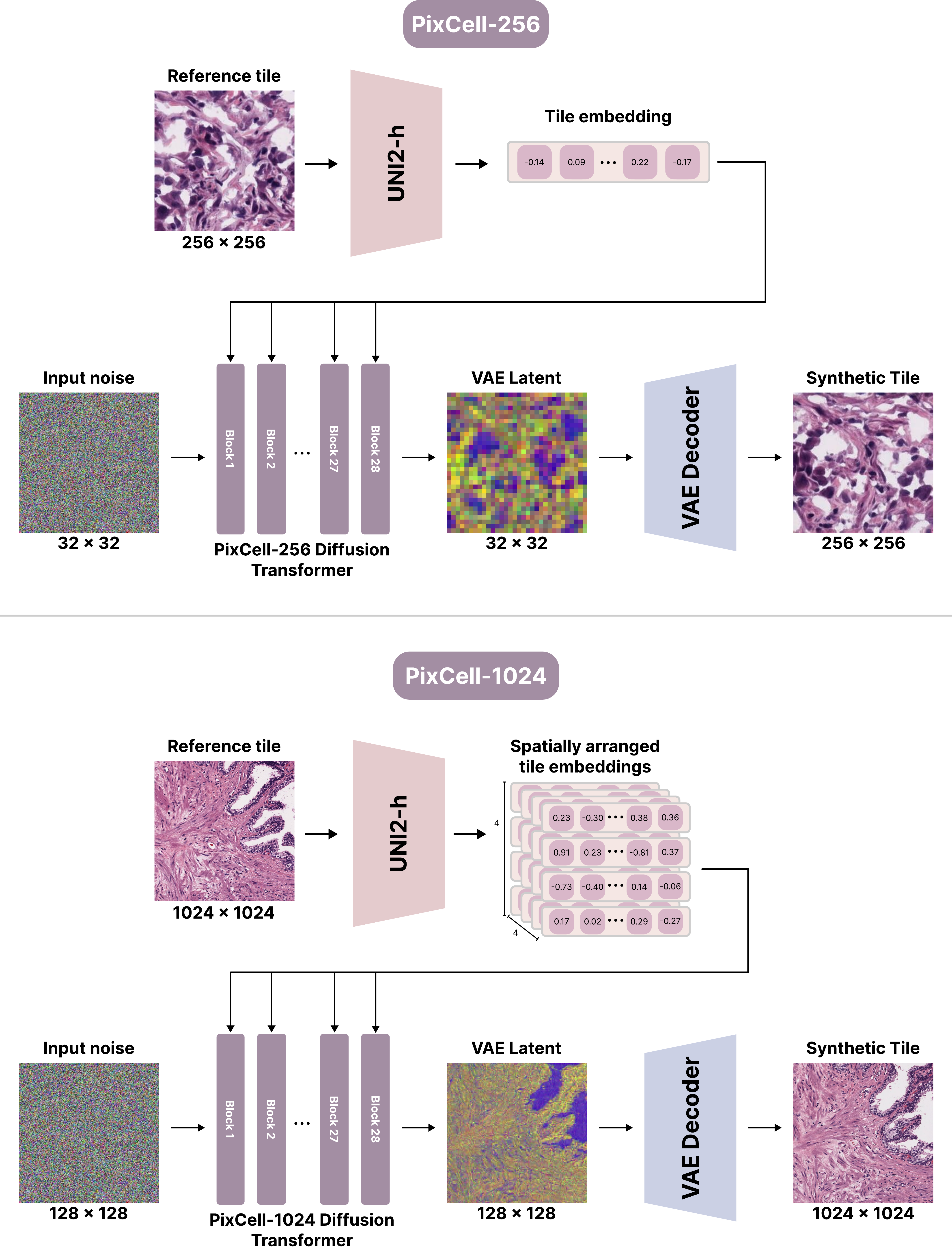}
    \caption{Overview of (a) the PixCell-256 model and (b) the PixCell-1024 model. Each model is conditioned on the UNI-2h embeddings that describe the given image.}
    \label{fig:pixcell_overview}
\end{figure}

PixCell is a Diffusion Transformer (DiT) model \cite{peebles2023scalable}, with its architecture adapted from the PixArt-$\sigma$ framework \cite{pixart-sigma}. We follow the LDM approach \cite{rombach2022high}, operating in the latent space of the pre-trained Variational Autoencoder (VAE) from Stable Diffusion 3 \cite{sd3}. The VAE first encodes the high-resolution histopathology images from PanCan-30M into compressed latent representations, on which we train the diffusion model. The VAE encoder has a downsampling factor of $\times 8$, compressing a $1024 \times 1024 \times 3$ image to a $128 \times 128 \times 16$ latent. We opted for the Stable Diffusion 3 VAE, despite the larger latent representation (16-dim in SD3 vs 4-dim in SDXL \cite{podellsdxl}), because of the better reconstruction quality in pathology images (average PSNR of 31.79 for SD3 vs 26.93 for SDXL \footnote{PSNR measured on 1000 random $256\times256$ tiles from PanCan-Test.}).

To guide PixCell during generation, we condition the model on image embeddings extracted using the self-supervised UNI-2h \cite{uni} encoder. In the natural image domain, generative foundation models are usually trained on very large datasets of paired images and text captions.  As there are no large-scale pathology image-caption datasets for training, we utilize SSL embeddings instead of text captions \cite{cvpr24_ldm}. By conditioning on SSL embeddings, we replace the text prompt given in a text-to-image inference with a reference image `prompt'. The embeddings are integrated into the DiT blocks via cross-attention.

\subsubsection{Progressive training}
Instead of directly training PixCell on the 30M $1024\times1024$ image patches, we employ a progressive, multi-stage training strategy with increasing image resolution. The coarse-to-fine approach is computationally efficient, allowing the model to learn local structures at lower resolutions (where iterations are faster) before learning to model the consistent, long-range tissue architecture required for the larger regions. This technique is also used for the PixArt family of models \cite{pixart-alpha, pixart-sigma} in the natural image domain.

We divide the training into three stages that utilize the same underlying PanCan-30M dataset of $1024\times1024$ pixel patches. For the lower-resolution training stages, multiple smaller patches are derived from each original high-resolution patch. To save compute time, we pre-extract VAE features for all $1024 \times 1024$ patches in our PanCan-30M dataset, using the pretrained SD-3 VAE \cite{sd3}, resulting in $128 \times 128 \times 16$ dimensional latents. Simultaneously, we extract the $16 \times 1536$ dimensional conditioning embedding using the pretrained UNI-2h \cite{uni} encoder, where each $1 \times 1536$  embedding embedding corresponds to a single $256\times256$ tile in the image. 

We also create a separate hold-out test set of 2.3 million image patches, which we call \textbf{PanCan-Test}. We utilize the pre-extracted VAE features and UNI-2h embeddings corresponding to the remaining 27.9 million image patches for all training stages of PixCell:

\paragraph{Stage 1 - Low-resolution training (PixCell-256):} We first train a base model on latent features  corresponding to $256 \times 256$ px images. For each $1024 \times 1024$ image's pre-extracted features, we generate 16 non-overlapping $32 \times 32 \times 16$ crops from its $128 \times 128 \times 16$ VAE latent. Each of these latent crops is then paired with the corresponding $1 \times 1536$ embedding from the original $16 \times 1536$ UNI-2h embeddings. This stage is trained for one epoch over the entire dataset, yielding the \textbf{PixCell-256} model (Figure~\ref{fig:pixcell_overview} (a)).

\paragraph{Stage 2 - Intermediate fine-tuning:} The weights from PixCell-256 are used to initialize the model for a second stage of fine-tuning, this time using 512×512 images. From each full-resolution VAE latent, we extract 4 non-overlapping $64 \times 64 \times 16$ crops. Each $512 \times 512$ equivalent latent crop is paired with its corresponding $4 \times 1536$ embeddings (representing a $2\times2$ block of tokens) from the original UNI-2h embeddings extracted. This stage serves to gradually adapt the model before the final high-resolution training.
 
\paragraph{Stage 3 - High-resolution fine-tuning (PixCell-1024):} Finally, the model from the $512\times512$ stage is further fine-tuned on latent representations corresponding to full $1024\times1024$ pixel images for one epoch. The final $1024 \times 1024$ training stage directly uses the full pre-extracted $128 \times 128 \times 16$ VAE features and the complete $16 \times 1536$ UNI-2h embeddings. This final stage produces the high-resolution \textbf{PixCell-1024} model (Figure~\ref{fig:pixcell_overview} (b)). 

Table~\ref{tab:training_details} summarizes key training parameters, such as the number of iterations and batch sizes, for each stage. Across all stages, we use the AdamW optimizer with a weight decay of $0.03$ and a constant learning rate of $2 \times 10^{-5}$. We trained our models on a cluster with 32 NVIDIA A100 GPUs. More details are given in Appendix~\ref{sec:appendix_training_details}.

\begin{table}[ht]
    \centering
    \caption{Training details of PixCell. We progressively scale the model training resolution and resources.}
    \input{tables/training_details}
    \label{tab:training_details}
\end{table}

\subsection{Data Augmentation}
\label{sec:method_data_aug}

PixCell generates each image by conditioning on a semantic embedding from UNI-2h, ensuring that the synthetic image shares semantic features with the real reference image. We use this principle to enhance downstream diagnostic models through two distinct data augmentation strategies: a vanilla approach and a more targeted, constraint-guided approach.

\subsubsection{Vanilla data augmentation}

For our primary augmentation strategy, we leverage the semantic consistency between the real and synthetic images. Following the approach of LRDM \cite{cvpr24_ldm}, we create a synthetic counterpart for each real image in the training set and assign it the same class label as the reference. This method, which we term vanilla augmentation, can effectively double the size of the training dataset, offering a wider variety of semantically equivalent examples to the classifier.

\subsubsection{Test-time constraint-guided augmentation}

We also implement a more targeted augmentation strategy that aligns the synthetic images more closely with the feature space of the downstream classifier. This helps produce augmentations for encoders that the diffusion model has not been trained with (Virchow2, Hoptimus-1). We use a test-time algorithm for diffusion models \cite{graikos2024fast}, applying an additional constraint during sampling that maximizes the similarity between the real and synthetic images within the embedding space of the target pathology foundation model. For each classification task, the same foundation model is used as the backbone for the linear probe classifier and for constrained sampling.

\subsection{Virtual Staining}
\label{sec:virtual_staining}

PixCell is trained to generate H\&E-stained histopathology images conditioned on UNI-2h embeddings. However, the UNI encoder has been trained on both H\&E and immunohistochemistry (IHC) images. Therefore, we begin by examining whether our model can generalize to produce images of unseen stains using different conditions.

To test this, we prompt PixCell-256 using an IHC image from the MIST-HER2 dataset \cite{li2023adaptive}. In the MIST dataset, H\&E and IHC slides are generated from pairs of serial tissue sections that are stained and scanned separately. Each whole slide image is partitioned into patches; pairs of  patches are approximately aligned.  We extract the UNI-2h embedding of the IHC image and use it to condition the PixCell model. The generated images in Appendix Figure~\ref{fig:pixcell_ihc_variations} follow the appearance and semantics of the reference IHC patch, showing that, although unseen during diffusion training, the model can infer results that would have been obtained with different stains.

IHC stains reveal the spatial distribution of specific molecular markers and are significantly more expensive to obtain than H\&E. Given that we can generate IHC images, we propose to utilize the pre-trained PixCell model to perform stain translation between H\&E and IHC images. Given a dataset of `roughly' paired H\&E and IHC images, such as MIST \cite{li2023adaptive}, we introduce a method to generate an IHC-stained image from a given H\&E image of the same tissue region. Using a generative model for virtual (re-)staining has the potential to enable further downstream analysis of H\&E-stained tissues, without requiring an expensive and likely unavailable IHC staining procedure.

To adapt PixCell for the virtual staining task, we first learn how to translate the UNI-2h embeddings that our model is conditioned on from the H\&E domain to the target IHC stain domain. Given (loosely) paired patches from a dataset, we can train a rectified flow \cite{liu2022flow} that transforms input embeddings in the H\&E domain to the target IHC domain embeddings. Then, by simply transforming the conditioning extracted from a reference H\&E tile to the appropriate IHC stain conditioning, we can utilize PixCell to generate a new image in the target IHC domain.

\paragraph{MIST dataset} The first dataset we use to evaluate our stain translation approach is the MIST dataset \cite{li2023adaptive}. MIST contains roughly paired H\&E and IHC tiles of size $1024 \times 1024$, for four different IHC stains: HER2, ER, PR and Ki67.

To learn the H\&E UNI to IHC UNI mapping, we extract loosely paired $256\times256$ crops from the original $1024 \times 1024$ images in MIST and train a rectified flow model \cite{liu2022flow}. Rectified flows learn to transform samples between two distributions, in this case, H\&E UNI embeddings and IHC UNI embeddings. To implement the rectified flow model, we use a simple residual MLP network \cite{li2024return}. For each pair of images, we extract the two corresponding UNI embeddings (H\&E and IHC) and train the rectified flow model to learn the velocity field from H\&E to IHC embeddings. 

To perform virtual staining, we start from the H\&E UNI embeddings extracted from the reference image and integrate the ODE defined by the trained rectified flow velocity to sample the corresponding IHC embeddings. We then feed these embeddings to PixCell to synthesize the IHC image. We utilize the PixCell-1024 model which generates images of the same size as the MIST dataset. An advantage of this conditioning-transformation approach is that we operate entirely on a patch level, which is desirable as most of the paired H\&E and IHC datasets lack pixel-perfect alignment. We could also directly fine-tune the model to perform the stain translation but that would require (i) a much larger dataset and (ii) perfectly aligned H\&E and IHC images.

\begin{figure}[ht]
    \centering
    \includegraphics[width=1\linewidth]{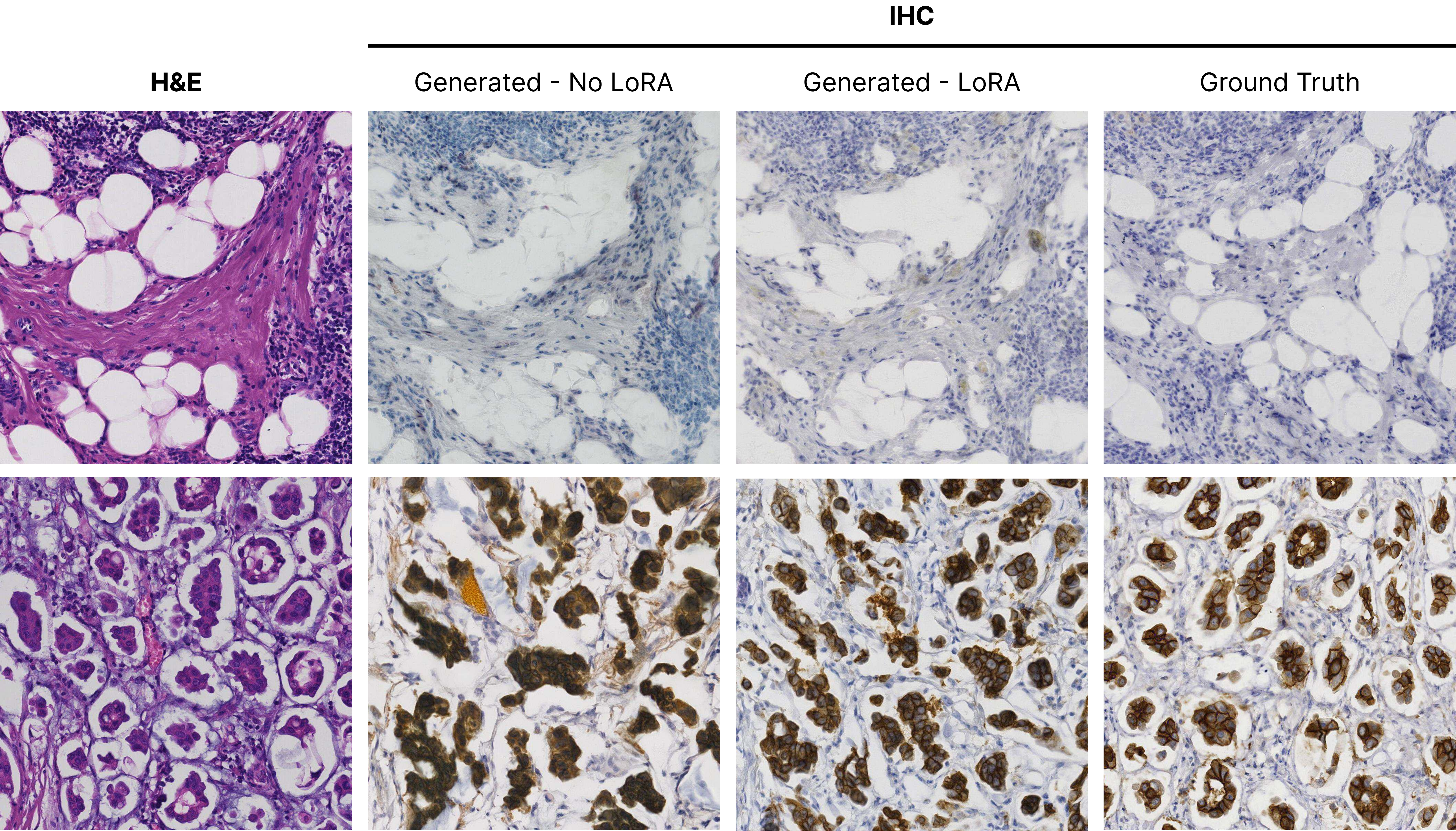}
    \caption{Results for H\&E \textrightarrow\ IHC stain translation using PixCell-1024 with and without LoRA fine-tuning. The quality of the generated images significantly improves with a lightweight low-rank adapter.}
    \label{fig:stain_translation_lora}
\end{figure}

In Figure~\ref{fig:stain_translation_lora} we present the results of this stain translation. We select H\&E images from the test set, extract their UNI embeddings, transform them to IHC embeddings with the trained rectified flow model, and generate corresponding IHC images. The resulting samples follow the overall semantics of the ground truth IHC images but noticeably lack in visual quality. Apart from not being trained on IHC data, we attribute the worse quality to the MIST dataset being at a slightly different resolution (0.4661 microns per pixel) than the dataset used for training (0.5 microns per pixel). 

To improve the quality of the generated images, we train a lightweight low-rank adapter (LoRA) \cite{hu2022lora} for the target dataset. Training a LoRA requires minimal computational resources compared to fine-tuning the entire diffusion model; we train the LoRA on the MIST images for approximately 10,000 iterations, on a single NVIDIA A5000 24GB GPU. The adapter massively improves the generated IHC image quality, without modifying any other component of the stain translation pipeline (Figure~\ref{fig:stain_translation_lora}).

\paragraph{HER2Match dataset} We also apply our virtual staining pipeline to the HER2Match dataset \cite{klockner2025gans}. The reported baselines were trained on $40\times$ magnification and $512\times512$ patches. We use the paired data to train the rectified flow that will translate H\&E embeddings to HER2 embeddings and utilize PixCell-1024, which we train on the same concentric patches at $20\times$ magnification, generating larger tissue regions. We apply the same training scheme for training the USIGAN \cite{peng2025usigan} baseline, which we find performs best among all models reported in the HER2Match paper.

\begin{figure}[t]
    \centering
    \includegraphics[width=1.0\linewidth]{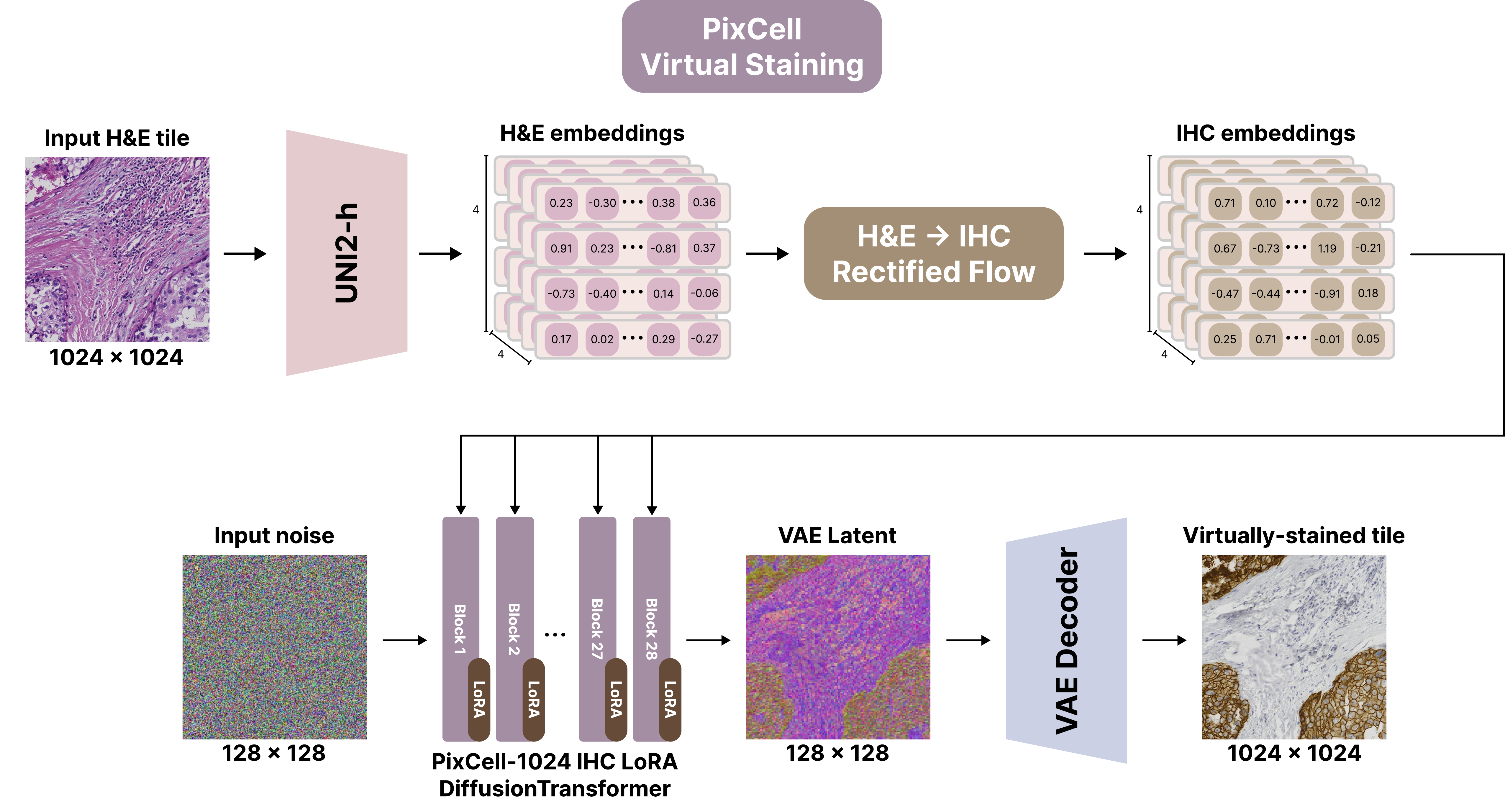}
    \caption{Overview of our approach for translating H\&E stained images to IHC using PixCell-1024. We train an additional H\&E-to-IHC model that transforms the UNI conditions from one domain to the other. Adding a low-rank adapter to the PixCell-1024 weights significantly improves the synthesized image quality with minimal additional costs.}
    \label{fig:pixcell_stain_transfer}
\end{figure}

%% file: tables/data_sources.tex
\begin{tabular}{ccc}
    \toprule
    Data Source          & \# WSI & \# Patches \\
    \midrule
    TCGA - diagnostic    & 11,766  & 9,192,388   \\
    TCGA - fresh frozen  & 18,310  & 3,698,529   \\
    CPTAC                & 6,124   & 2,200,109   \\
    SBU                  & 5,698   & 6,689,518   \\
    GTex normal slides   & 25,713  & 7,896,145   \\
    Other public sources & 1,573   & 1,143,288   \\
    \midrule
    Total                & 69,184  & 30,819,977 \\
    \bottomrule
\end{tabular}

%% file: tables/training_details.tex
\begin{tabular}{cccccc}
    \toprule
    Resolution & \# Images & Training steps & Batch size & \# GPUs & Conditioning \\ 
    \midrule
    $256 \times 256$ & 480 Mil & 120,000 & 4096 & 16 & $1 \times 1536$ \\
    $512 \times 512$ & 120 Mil & 60,000 & 1536 & 24 & $4 \times 1536$ \\
    $1024 \times 1024$ & 30 Mil & 80,000 & 384 & 32 & $16 \times 1536$ \\ 
    \bottomrule
\end{tabular}

%% file: Sections/experimental_setup.tex
\section{Experimental Setup}

\subsection{Image quality}

We evaluate image generation quality at $256 \times 256$ resolution using metrics that capture both generic and pathology-specific aspects of realism. We compare PixCell-256 against prior diffusion-based methods - LRDM \cite{cvpr24_ldm} and ZoomLDM \cite{zoomldm}. We sample 10,000 images to perform evaluation, following PathLDM \cite{pathldm}. 

\noindent \textbf{Evaluation datasets:}
We assess PixCell on three complementary datasets that test distinct aspects of generalization:

\begin{itemize}
    \item PanCancer-Test (within-distribution): We create a separate hold-out test set of 2.3 million image patches ($1024 \times 1024$ px at $20 \times$ magnification) sampled from the same data sources used for training. The data was split at the patch-level, not slide-level.
    \item SPIDER (cross-source): We use the publicly available SPIDER dataset \cite{nechaev2025spider}, which contains images from four organ types -- Breast, Colorectal, Skin and Thorax, collected at a different imaging center. We use 339,330 $256 \times 256$ patches at $20 \times$ magnification.
    \item Non-H\&E (stain robustness): We collect 6971 WSIs from the HistAI-mixed \cite{histai} dataset that were prepared with alternative staining protocols, and extract 3.4 Million patches at $20 \times$ magnification.
\end{itemize}

Images from SPIDER and HistAI datasets were not included in PixCell training.

\noindent \textbf{Metrics:}

\begin{itemize}
    \item Fréchet Inception Distance (FID): measures the distance between feature distributions of real and synthetic images using Inception-v3 activations \cite{heusel2017gans, szegedy2016rethinking}. Lower value indicates higher visual fidelity.
    \item Fréchet Hoptimus/Virchow Distance (FHD/FVD): computed identically to FID but in the feature space of the Hoptimus-1 / Virchow-2 \cite{hoptimus1, virchow2} foundation models, capturing pathology-specific structures. Lower value denotes closer alignment to real tissue morphology.
    \item Embedding similarity: computes cosine similarity between real and synthetic image embeddings. We use three pathology foundation models: Hoptimus-1, KEEP \cite{keep} and Virchow-2. Higher similarity indicates better preservation of pathology features.
\end{itemize}

For $1024\times1024$ high-resolution images in Appendix \ref{sec:high_res_supp}, we adopt an evaluation strategy consistent with LRDM \cite{cvpr24_ldm} and MultiDiffusion  \cite{bar2023multidiffusion}. This strategy involves two primary perceptual metrics:

\begin{itemize}
    \item Crop FID:  We compare the distribution of $256\times256$ random crops we extract from the synthesized images against the distribution of real $256\times256$ patches using the vanilla Inception-v3 network \cite{szegedy2016rethinking}. 
    \item CLIP FID: We measure the FID directly between the synthesized images generated by PixCell-1024 and real $1024\times1024$ images. For this, we extract features using a pretrained CLIP ViT-B/32 model \cite{radford2021learning, KynkaanniemiKAA23}, which is better-suited to evaluating larger images ($>224\times224$ px). 
\end{itemize}

We use the Clean-FID \cite{parmar2021cleanfid} package to measure all Fréchet distances. We choose KEEP, Virchow-2 and Hoptimus-1 because they represent state-of-the-art vision-language and SSL encoders \cite{thunder} trained on diverse histopathology data.

\noindent\textbf{Pathologist Evaluation - Image quality:} We generated $4096 \times 4096$ synthetic regions using PixCell and ZoomLDM. We then presented pairs of real and synthetic images to two board-certified pathologists in a blinded fashion, with no information about the source model. For each synthetic region, the pathologist scored five criteria: tissue architecture, cell morphology, sharpness, color fidelity, and realism (absence of hallucination artifacts). All criteria were rated on a 1-3 ordinal scale, where higher values indicate better image quality (sharper images, lesser hallucinations etc). We collected N=36 samples for BRCA and N=50 for PRAD cancer types.

\noindent\textbf{Pathologist Evaluation - subtype classification:} An expert pathologist performed lobular vs ductal carcinoma subtyping on synthetic BRCA regions (N=18 WSIs). For each WSI, we sampled eight synthetic and eight real regions from the same slide (all regions at $4096 \times 4096$ resolution). The pathologist evaluated the synthetic and real sets separately, producing two subtype predictions per WSI. We then measured subtyping accuracy with synthetic and real predictions independently.

\noindent\textbf{Statistical analysis:} For Fréchet distances, we estimate $95 \%$ confidence intervals across five random seeds, computed as $1.96 \times$ the standard error of the mean. Error bars in Fig. \ref{fig:pixcell_image_quality}  correspond to these intervals, and differences are considered meaningful when intervals do not overlap. For the subtyping accuracy plot, the error bars are 95\% Wilson score confidence intervals for a binomial proportion.

\subsection{Data augmentation}
\label{sec:setup_data_aug}

\begin{table}[ht]
\centering
\caption{Datasets used for augmentation}
\input{tables/aug_datasets}
\label{tab:aug_datasets}
\end{table}

\noindent\textbf{Datasets:} We use four public patch-level classification datasets as shown in Table \ref{tab:aug_datasets}.

\noindent \textbf{Linear probing:} Our evaluation protocol for linear probing strictly follows the Thunder framework \cite{thunder}. For each downstream task, we train a linear probe on a frozen encoder backbone for 200 epochs. We use the official train, validation, and test splits provided by Thunder.

To determine the optimal settings, we conduct a hyperparameter grid search for each combination of dataset and encoder, and select the final model based on the highest achieved accuracy on the validation set. The search space includes:

\begin{itemize}
    \item Learning rate: $\{ 10^{-3}, 10^{-4}, 10^{-5} \}$
    \item Weight decay: $\{0, 10^{-3}, 10^{-4}\}$
    \item Augmentation strategies: $\{\texttt{Vanilla Aug}, \texttt{Test-time Aug}\}$
    \item Diffusion guidance scale: $\{1,2,3,4\}$
\end{itemize}

\noindent \textbf{Statistical Analysis:} To ensure the statistical significance of our results, we generate $95\%$ confidence intervals (CIs) and perform one-sided p-tests. These are calculated via bootstrapping with 10,000 samples drawn with replacement.

\subsection{Synthetic SSL}

\begin{table}[ht]
    \centering
    \caption{Overview of downstream tasks/datasets used in our SSL experiments following Thunder benchmark~\cite{thunder}.}
    \input{tables/ssl_downstream_datasets}
    \label{tab:ssl_downstream_datasets}
\end{table}

\noindent \textbf{SSL training datasets:} We create two large-scale datasets for SSL training using the TRIDENT framework \cite{trident}. We extract 10 million $256 \times 256$ patches from 11,000 TCGA WSIs \cite{weinstein2013cancer} at $20 \times$ magnification (TCGA-Real). We then extract UNI-2h embeddings from these patches, and create a synthetic counterpart using PixCell-256 (TCGA-Synthetic). Similarly, we create HistAI-Synthetic using 10 million patches from 40,000 slides from the HistAI WSIs \cite{histai}.

\noindent \textbf{Model training:} We train Dinov2 ViT-Base encoders for  (1) a baseline model trained  on TCGA-Real, (2) a model trained  on TCGA-Synthetic to test for synthetic data replacement ability, and (3) a model trained  on a combined dataset of TCGA-Real and HistAI-Synthetic to test for privacy-preservation data scaling. For all configurations, we train the encoder from scratch for 100,000 iterations.

\noindent \textbf{Evaluation:} We evaluate the resulting frozen SSL encoders using the Thunder benchmark \cite{thunder}. We measure performance on 9 downstream patch-classification tasks using the k Nearest Neighbors (k-NN) approach. This training-free evaluation protocol ensures that the encoders are assessed based purely on the representations learned from their respective training sets. See Table \ref{tab:ssl_downstream_datasets} for more details on the datasets used.  

\subsection{Virtual Staining}

\paragraph{Training}
We train the LoRA for 10 epochs on each IHC stain dataset, using a batch size of 4 and a learning rate of $1\times10^{-4}$. We also drop the condition with probability $10\%$ to allow for classifier-free guidance \cite{ho2022classifier} during sampling. We add the low-rank adapters to the cross-attention layers of the diffusion transformer. We train the rectified flow Residual MLPs for 500 epochs for each IHC stain, using a batch size of 512 and a learning rate of $1\times10^{-4}$.

\paragraph{Inference}
During inference, we first extract the H\&E embeddings from the given image with UNI-2h and then transform them to the target IHC embeddings using the trained rectified flow MLP with the Euler ODE sampler for 100 steps. To synthesize a new virtually-stained image, we condition the diffusion transformer + LoRa model with the transformed embeddings, and apply classifier

\paragraph{Diagnostic evaluation} For the diagnostic evaluation we utilize DeepLIIF \cite{deepliif} to automatically segment stained and non-stained cells in real and synthesized IHC images. Then, for each specific stains we follow a specific guideline to determine an appropriate label. For ER, we label images as positive if 10\% or more of the tumor cell nuclei are stained, low positive if the percentage is between 1 and 10\%, and negative if it is less than 1\%. For PR, images are labeled as positive if more than 1\% of tumor cell nuclei are stained and negative otherwise. For Ki67, images are labeled as positive if more than 30\% of tumor cell nuclei are stained, negative if less than 5\% are stained, and inconclusive otherwise. When evaluating with an expert pathologist, we created an online website where the expert could view the H\&E and IHC images and provide one of the previous labels for the IHC image.

%% file: tables/aug_datasets.tex
\begin{tabular}{cccccc}
\toprule
Dataset   & Organ      & Num. of classes & Image size         & Magnification          & Num. of images \\ 
\midrule
BACH \cite{bach}          & Breast     & 4          & $1536 \times 2048$ & $20 \times$  & 408       \\
BRACS \cite{bracs}        & Breast     & 7          & Variable           & $40 \times$  & 4539      \\
Break-His \cite{breakhis} & Breast     & 8          & $700 \times 460$   & $40 \times $ & 1995      \\
MHIST \cite{mhist}        & CRC & 2          & $224 \times 224$   & $5 \times$   & 3152      \\ 
\bottomrule
\end{tabular}

%% file: tables/ssl_downstream_datasets.tex
\begin{tabular}{c|ccccc}
    \toprule
    Dataset &  Num. of classes & Organ & Image size & Magnification & Num. of images \\
    \midrule
    BRACS~\cite{bracs} & 7 & Breast & Variable &  40$\times$ & 4539 \\
    ccRCC~\cite{brummer2023computational} & 3 & Renal & 300 $\times$ 300 & 40$\times$ & 52173 \\
    NCT-CRC~\cite{nct-crc} & 9 & CRC & 224 $\times$ 224 & 20$\times$ & 107180 \\
    SPIDER breast~\cite{nechaev2025spider} & 18 & Breast & 224 $\times$ 224 & 20$\times$ & 92892 \\
    SPIDER colorectal ~\cite{nechaev2025spider} & 13 & CRC & 224 $\times$ 224 & 20$\times$ & 77182 \\
    SPIDER skin~\cite{nechaev2025spider} & 24 & Skin & 224 $\times$ 224 & 20$\times$ & 159854 \\
    SPIDER thorax~\cite{nechaev2025spider} & 14 &  thorax & 224 $\times$ 224 & 20$\times$ & 78307 \\
    TCGA CRC msi ~\cite{kather2020histological} & 2 & CRC & 512 $\times$ 512 & 20$\times$ & 52713 \\
    TCGA TILs~\cite{komura2020histology} & 2 & Multi & 100 $\times$ 100 & 20$\times$ & 304097 \\
    \bottomrule
\end{tabular}

%% file: Sections/related_work.tex
\section{Related Work}
\subsection{Self-Supervised Learning in Computational Pathology}
The digitization of histology slides into Whole Slide Images (WSIs) has revolutionized pathology, providing a vast amount of visual data for cancer diagnosis and biomedical research. These gigapixel-scale images capture complex tissue architecture and cellular morphology, offering substantial potential for computational analysis. However, the sheer size of WSIs necessitates processing them at the patch level. A significant challenge arises from the lack of precise, patch-level annotations, making traditional supervised learning approaches difficult. This annotation bottleneck has driven the significant traction of foundation models in digital pathology, particularly self-supervised learning (SSL) at the tile-level~\cite{uni, virchow1, phiconv2, hibou, kaiko, gigapath}, which allows models to learn meaningful representations from large unlabeled datasets.

SSL enables models to learn generalizable features from the inherent structure of the data itself through pretext tasks. Joint-embedding SSL (JE-SSL) methods, a prominent subclass, aim to learn effective representations by promoting alignment between embeddings of augmented views of the same image while ensuring diversity across the entire dataset. Self-distillation variants within JE-SSL, such as the DINOv2 framework, have demonstrated considerable success in natural image understanding and have been widely adopted in computational pathology (CPath)~\cite{dino, ibot, dinov2}. These tile-based SSL models, including notable examples like UNI~\cite{uni}, Virchow~\cite{virchow1}, Prov-GigaPath~\cite{gigapath}, and Phikon~\cite{phiconv2}, are trained on increasingly large and diverse datasets. Early efforts often utilized public repositories like The Cancer Genome Atlas (TCGA), while more recent models leverage extensive proprietary datasets, encompassing millions of slides and billions of tiles. This data often spans multiple institutions, disease states, tissues of origin, and includes a growing variety of staining techniques beyond hematoxylin and eosin (H\&E), such as immunohistochemical (IHC) and chemical stains, as well as data from multiple image magnifications.

Beyond SSL, other discriminative models leverage contrastive learning with paired image-text datasets \cite{plip, quilt, conch, musk, sun2024pathgen}. Addressing the gigapixel scale of WSIs, slide-level foundation models have also been developed \cite{titan, gigapath}. These operate on larger image regions to capture slide-level context, often by building upon embeddings generated by patch-level encoders (many of which are trained via SSL) in a hierarchical fashion \cite{tangle, gecko}. The quality of these patch-level SSL encoders is thus a critical performance gateway for such hierarchical systems.

\subsection{Diffusion models}

Diffusion models \cite{ho2020denoising} have established themselves as the main approach for building large-scale generative foundation models. Significant advances in training paradigms \cite{improved_ddpm, song2021scorebased, lipman2023flow}, model architectures \cite{ldm, peebles2023scalable} , guidance techniques \cite{ho2022classifier, zhang2023adding, graikos2023conditional}, and sampling strategies \cite{ddim, dpm} have enabled  exceptional capabilities in synthesizing high-fidelity, diverse images. Latent Diffusion Models (LDMs)~\cite{rombach2022high}, specifically, reduce the computational cost by compressing the images with a learned encoder-decoder pair \cite{pathvae}.  As also observed by early works on diffusion models \cite{dhariwal2021diffusion}, accurate conditioning plays a critical role in achieving high-quality image generation.

Examining popular diffusion foundation models in natural images (SD-XL \cite{podellsdxl}, SD-3 \cite{sd3}, PixArt \cite{pixart-alpha, pixart-sigma}), most utilize the LDM approach with a pre-trained encoder and decoder, and train on datasets of billions of image-text pairs to train a text-conditioned diffusion model. Beyond data augmentation \cite{TrabuccoDGS24, AziziKS0F23, gen-sis}, the powerful image priors encoded by such foundational generative models have been effectively utilized for a wide range of dense prediction tasks, including image segmentation \cite{diffuse_attend} and depth estimation \cite{Ke_2024_CVPR}. A primary factor contributing to the success of diffusion models in natural imaging is the availability of vast, curated collections of image-text pairs. To illustrate, Stable Diffusion 1.4 \cite{ldm} was trained on LAION-5B \cite{schuhmann2022laion}, a dataset comprising 5.85 billion such pairs. In comparison, datasets of this scale and nature are not readily available for histopathology. 

Despite these pronounced differences, diffusion models have been explored for digital pathology \cite{moghadam2023morphology,muller2023multimodal,pathldm,aversa2023diffinfinite}. Most are limited to training models on small-scale datasets, partly due to the lack of annotations for training a conditional diffusion model on large datasets. Notably, learned representation-guided diffusion models (LRDMs) \cite{cvpr24_ldm} provide an alternative by training the generative model conditioned on an embedding produced by a self-supervised encoder. This approach offers a path to scaling up diffusion models in the histopathology domain.

%% file: Sections/conclusion.tex
\section{Conclusion}
We developed PixCell, the first generative foundation model for histopathology images. Trained on PanCan-30M -- an extensive dataset derived from 69,184 H\&E-stained whole-slide images covering a comprehensive range of cancer types, PixCell demonstrates state-of-the-art image generation quality. It produces high-fidelity and diverse samples that can effectively serve as a replacement for real images in training self-supervised learning models. Beyond improving downstream performance through synthetic data augmentation, we show that PixCell can generate large synthetic counterparts of entire cohorts (e.g., TCGA, HistAI) that act as drop-in substitutes for real slides, enabling multi-institutional model training in settings where direct data sharing is not feasible. We explored PixCell's capability for controllable generation by conditioning on cell segmentation maps and performing synthetic data augmentation for downstream tasks. 
Finally, we demonstrated PixCell’s adaptability by showing that it generalizes to different staining techniques and can serve as a backbone for virtual IHC staining. Given an H\&E image, PixCell can synthesize corresponding IHC images that support quantitative biomarker analysis. With appropriate domain-specific validation, this virtual IHC capability has substantial potential for translational research studies.

We have publicly released the PixCell model weights and associated code to foster future research and applications in computational pathology. Our work with PixCell highlights the potential for diffusion-based generative models to serve as true foundational models within the histopathology domain, offering capabilities that extend beyond traditional discriminative approaches. We hope that this contribution will inspire the broader research community to further utilize these models for innovative applications and novel downstream tasks.

\section*{Acknowledgments}
This research was partially supported by National Institutes of Health (NIH) and National Cancer Institute (NCI) grants
1R21CA258493-01A1, 5U24CA215109, UH3CA225021, U24CA180924, National Science Foundation (NSF) grants IIS-2123920, IIS-2212046, as well as  Stony Brook Profund 2022 seed funding and private support from Bob Beals and Betsy Barton. This research used resources of the Argonne Leadership Computing Facility, a U.S. Department of Energy (DOE) Office of Science user facility at Argonne National Laboratory and is based on research supported by the U.S. DOE Office of Science-Advanced Scientific Computing Research Program, under Contract No. DE-AC02-06CH11357.

%% file: Sections/appendix.tex
\clearpage
\appendix

\section{Appendix}

\subsection{Supplementary file}

We include a supplementary \verb|.zip| file along with the main paper. The supplementary file contains several synthetic images (at 256 / 1024 / 4096 px resolution) along with the complete pathologist evaluation data. Once the zip file is extracted, the synthetic images can be visualized by opening the HTML files in  \verb|pixcell_gen_images| directory.

\subsection{High-resolution image generation}
\label{sec:high_res_supp}

\subsubsection{PixCell-1024}

\begin{table}[ht]
    \centering
    \caption{Image quality of $1024\times1024$ patches and generation time on TCGA-BRCA. PixCell-1024  outperforms all previous baselines in both sampling time and generation quality.}
    \input{tables/fid_1024}
    \label{tab:fid_1024}
\end{table}

For $1024 \times 1024$ generation, we compare PixCell-1024 on the TCGA-BRCA dataset against prior methods: LRDM \cite{cvpr24_ldm}, $\infty$-Brush \cite{inftybrush}, and ZoomLDM \cite{zoomldm}. Our assessment considers image quality, using Crop FID and CLIP FID, as well as generation efficiency, measured in generation time per image (seconds). 

As shown in Table~\ref{tab:fid_1024}, PixCell-1024 demonstrates SoTA performance across all evaluated metrics. For image quality, it achieves a Crop FID of 7.92 and a CLIP FID of 0.68, substantially lower than prior work. This signifies that images generated with PixCell-1024 more closely resemble the feature distribution of real images, compared to those from baseline methods. Furthermore, PixCell-1024 achieves the best generation speed, producing a $1024\times1024$ image in just 2.5 seconds, significantly faster than its counterparts. These findings, coupled with the strong results shown by PixCell-256, highlight the scalability of our PixCell architecture and the effectiveness of the progressive, pan-cancer training strategy from $256\times256$ up to $1024\times1024$ resolution.

\begin{figure}[ht]
    \centering
    \includegraphics[width=1.\linewidth]{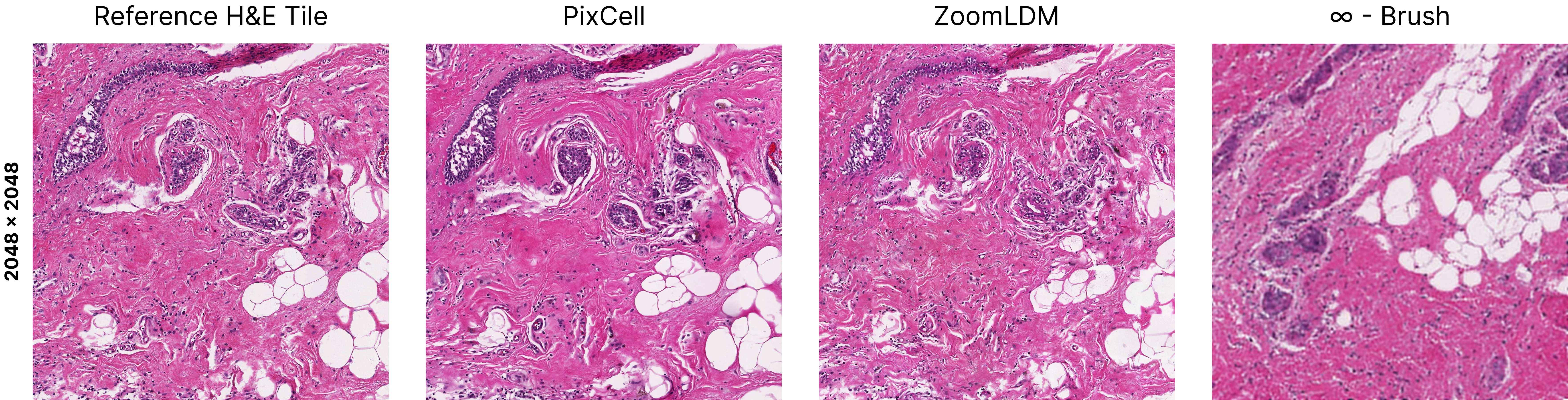}
    \caption{We compare a $2048 \times 2048$ image generated by previous methods - $\infty$-Brush \cite{inftybrush} and ZoomLDM \cite{zoomldm}. $\infty$-Brush produces blurry details and ZoomLDM generates less realistic structures.}
    \label{fig:snake}
\end{figure}

\begin{figure}[ht]
    \centering
    \includegraphics[width=1.\linewidth]{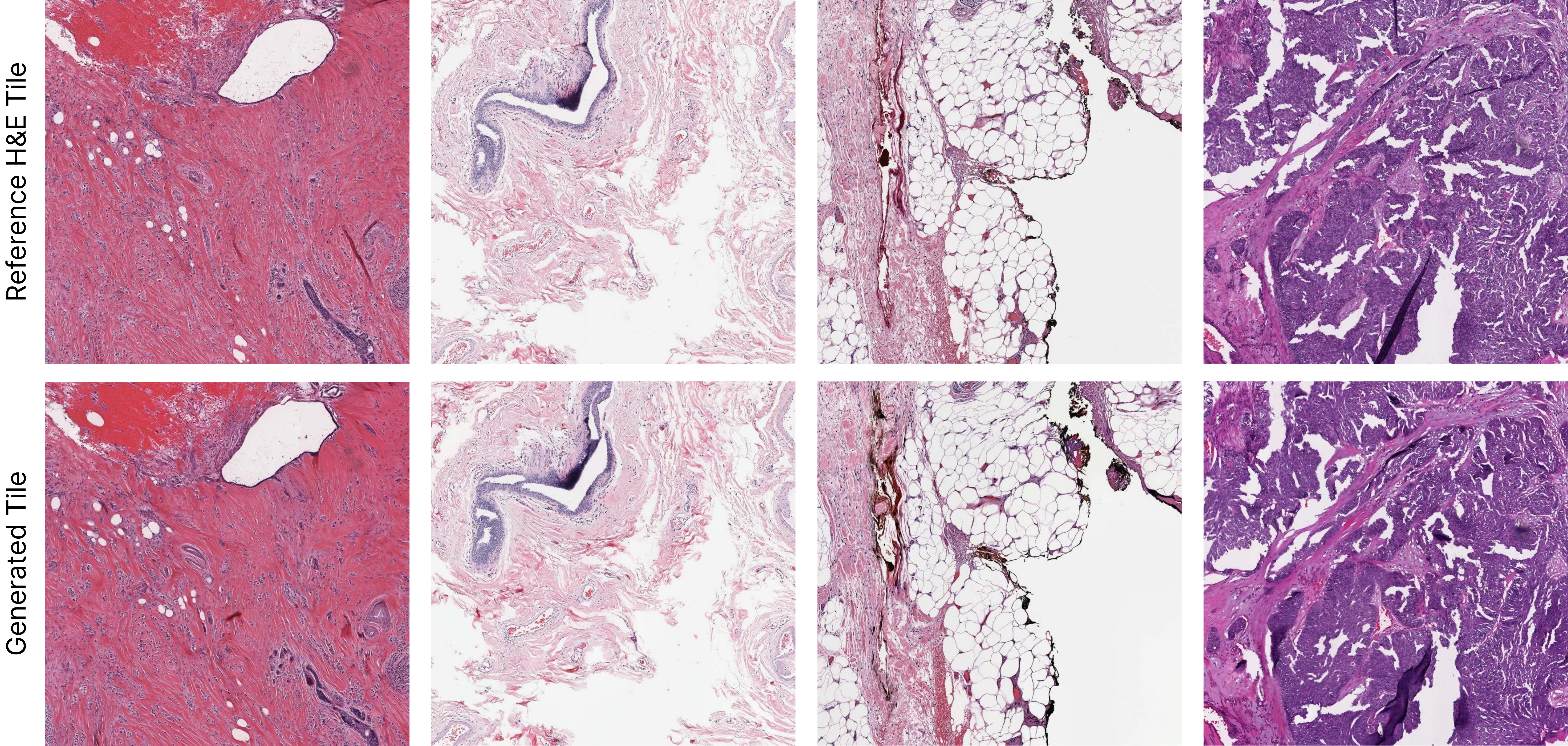}
    \caption{$4096 \times 4096$ images synthesized by PixCell-1024 using the patch-wise generation algorithm introduced in LRDM \cite{cvpr24_ldm}. Best viewed zoomed-in.}
    \label{fig:4k_images}
\end{figure}

\subsubsection{\texorpdfstring{$2048 \times 2048$}{2048x2048} and beyond}
To demonstrate PixCell's applicability for synthesizing huge high-resolution images ($2048 \times 2048$ and beyond), we adapt the patch-wise generation algorithm introduced in LRDM \cite{cvpr24_ldm}. While the original LRDM approach used a $256 \times 256$ resolution base model applied in a tiled fashion, we extend this strategy by employing our PixCell-1024 model as the patch-wise generator. By applying this algorithm, we successfully synthesize coherent 4096 $\times$ 4096 pixel histopathology images. As shown in Figure \ref{fig:4k_images},  the synthesized ultra-high-resolution images effectively maintain both the global tissue architecture and the fine-grained local cellular details of the reference image. Furthermore, we compare a $2048 \times 2048$ image generated by PixCell with prior work in Figure \ref{fig:snake}. This result showcases the robustness of PixCell's learned priors and its scalability to large-image synthesis tasks.

\subsection{Classifier-free guidance scale}
\begin{table}[ht]
    \centering
    \caption{Grid search to find the optimal classifier-free guidance scale $w$ based on the FID score.}
    \input{tables/cfg_scale}
    \label{tab:fid_cfg_search}
\end{table}

To determine the optimal classifier-free guidance scale hyperparameter, denoted as $w$, we perform a grid search across the guidance values $w \in \{1, 1.2, 1.4, 1.6, 1.8, 2, 3, 5\}$. For this search, we generate images conditioned on randomly selected UNI-2 embeddings from the training set and measure FID against images from the PanCan-Test set. Table~\ref{tab:fid_cfg_search} indicates that the best image quality (lowest FID) is obtained for $w=2$ for PixCell-256 and $w=1.2$ for PixCell-1024.

The optimal guidance scale is noticeably lower than the scales regularly used in text-to-image diffusion models for natural images (typically around 6). We attribute this to our choice of conditioning; whereas a single text caption can characterize different images (wider distribution), a UNI embedding corresponds to a much smaller set of images, i.e., the distribution of images the model learns to sample from given a UNI embedding is more concentrated. We use the optimal guidance values for all image generation tasks reported in this paper, unless stated otherwise.

\subsection{Data Augmentation}
\label{sec:supp_data_aug}

\begin{table}[ht]
\centering
\caption{Image encoders can overfit on small datasets, synthetic data boosts performance.}
\input{tables/data_augmentation_knn}
\label{tab:data_aug_knn}
\end{table}

Foundation models like UNI-2, Virchow-2, and Hoptimus-1 extract semantically rich features from image patches. However, downstream datasets often contain a limited number of training samples ($N_{\text{BACH}}=218, N_{\text{BRACS}}=3657, N_{\text{Break-his}}=936$ and $N_{\text{mHist}}=1743$ respectively), which presents a bottleneck. A training-free approach like k-Nearest Neighbors (k-NN) can be highly sensitive to this data scarcity.

As seen in Table \ref{tab:data_aug_knn}, the k-NN classifier's performance is significantly lower than that of a trained linear probe in every configuration. For example, on the BACH dataset with the U-2 encoder, the k-NN F1 score is 79.4, whereas the linear probe achieves 87.0. A linear classifier however, is far more effective at learning a robust decision boundary from the limited training data. The main section \ref{sec:results_aug} further shows how this linear probe's performance is substantially boosted by augmenting the small training set with PixCell-generated images.

\begin{table}[ht]
    \centering
    \caption{Performance of SSL encoders trained on multiple combinations of real and synthetic TCGA data}
    \input{tables/knn_ssl_supp}
    \label{tab:knn_ssl_supp}
\end{table}

\subsection{Synthetic SSL Scaling}
\label{sec:syn_ssl_scaling}

To further investigate the scaling properties of synthetic data in SSL pretraining, we trained additional DINO-v2 models on different combinations of real and synthetic data (Table \ref{tab:knn_ssl_supp}). We observe that simply augmenting the full real dataset with its complete synthetic copy (\texttt{TCGA Real 100\% + Syn 100\%}) does not improve performance over \texttt{TCGA Real 100\%} baseline ($76.76 \%$ vs $77.89 \%$). This finding confirms that incorporating new, unseen data distributions from other sources (like HistAI), even as synthetic samples, can improve model generalization.

We also simulate a data-constrained scenario. A model trained on only real images from half the TCGA slides (\texttt{TCGA Real 50\%}) achieves an average accuracy of $76.22 \%$. However, a model trained on real images from half the slides and synthetic from the other half (\texttt{TCGA Real 50\% + Syn 50 \%}) achieves an accuracy of $76.87 \%$. This demonstrates that synthetic data can effectively stand in for the missing real data. As expected, this performance is still upper-bounded by the model trained on 100\% real data, but it shows a viable path for recovering performance when a complete real dataset is not available.

\subsection{Agreement between expert pathologist and automated scoring}
We evaluate the agreement between the automated scoring, using DeepLIIF \cite{deepliif}, and an expert pathologist's scoring. For the three IHC stains, we measure the accuracy of the DeepLIIF predictions, using the expert pathologist's evaluations as ground truth, as well as Cohen's $\kappa$, and present the results in Table~\ref{tab:ihc_agreement}. We find that, overall, DeepLIIF performs well as an automated scoring method, making it a useful tool in evaluating virtual staining pipelines.

\begin{table}[ht]
    \centering
    \caption{Agreement between pathologist and automated scoring.}
    \begin{tabular}{lcccccc}
        \toprule
        & \multicolumn{2}{c}{ER} & \multicolumn{2}{c}{PR} & \multicolumn{2}{c}{Ki67} \\
        \cmidrule(lr){2-3} \cmidrule(lr){4-5} \cmidrule(lr){6-7}
        Images & Acc & $\boldsymbol{\kappa}$ & Acc & $\boldsymbol{\kappa}$ & Acc & $\boldsymbol{\kappa}$ \\
        \midrule
        Real              & 0.714 & 0.762 & 0.767 & 0.516 & 0.500 & 0.554 \\
        PixCell-generated & 0.667 & 0.547 & 0.552 & 0.253 & 0.381 & 0.382 \\
        USIGAN-generated  & 0.552 & 0.417 & 0.552 & 0.226 & 0.526 & 0.390 \\
        All               & 0.643 & 0.593 & 0.625 & 0.342 & 0.467 & 0.504 \\
        \bottomrule
    \end{tabular}
    \label{tab:ihc_agreement}
\end{table}

\subsection{Controllable generation}
\label{sec:controllable_gen}
\begin{figure}[t]
    \centering
    \includegraphics[width=0.85\linewidth]{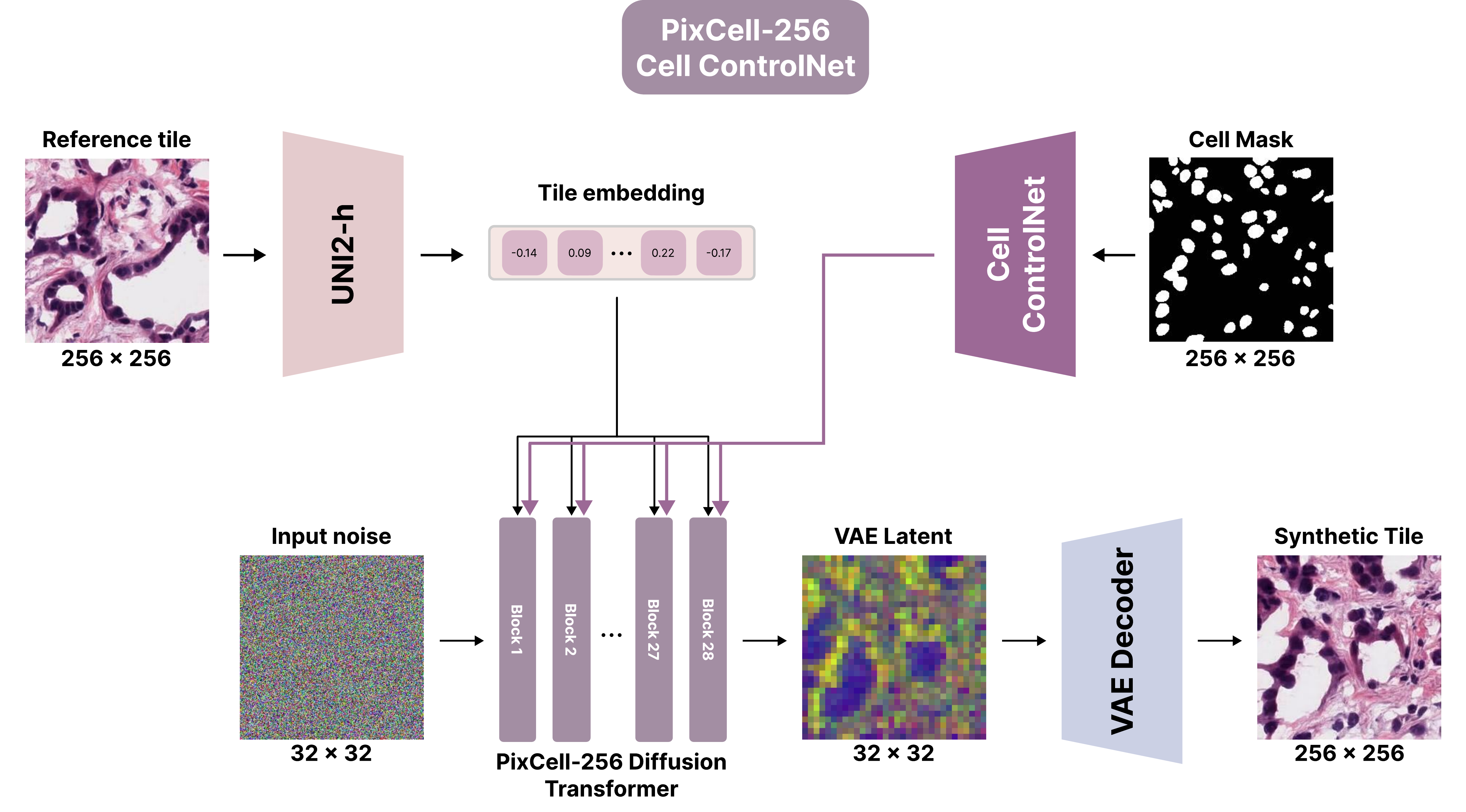}
    \caption{PixCell-256 with a cell mask ControlNet. The two sources of conditioning allow for fire control over the generated image: the UNI embedding dictates the style of the generated image while the cell mask guides the cell layout.}
    \label{fig:pixcell_controlnet}
\end{figure}

PixCell generates images conditioned on an embedding from a self-supervised encoder. Within this framework, we can synthesize variations of a given image but lack further control over the synthesized image content. To introduce the ability to controllably generate images, we employ ControlNets \cite{zhang2023adding}, which add image-level control to pre-trained diffusion models.

We train a ControlNet for PixCell-256 to guide generation with a cell layout mask. To construct a training dataset, we utilize a pre-trained CellViT-SAM-H \cite{horst2024cellvit} model trained on 0.5 microns per pixel images. We extract cell masks from 10,000 images of all cancer types from PanCan-30M and train the ControlNet with image, UNI embedding, and mask triplets. The diffusion transformer ControlNet copies each layer of the base transformer, adding an intermediate output linear layer that is zero-initialized to combine the base transformer and ControlNet transformer features \cite{chen2024pixart}. An overview of the ControlNet approach is provided in Figure~\ref{fig:pixcell_controlnet}.

\subsubsection{ControlNet training}
To generate $256\times256$ images conditioned on cell segmentation masks, we employed a pre-trained CellViT-SAM-H \cite{horst2024cellvit} model trained on $20\times$ pathology images. We randomly sampled 10,000 images from PanCan-30M and their UNI-2h embeddings, extracted binary cell masks using CellViT, and trained the ControlNet model for 25,000 iterations, using a batch size of 4 and the AdamW optimizer with a learning rate of $1\times10^{-5}$.

\begin{figure}[t]
    \centering
    \includegraphics[width=1\linewidth]{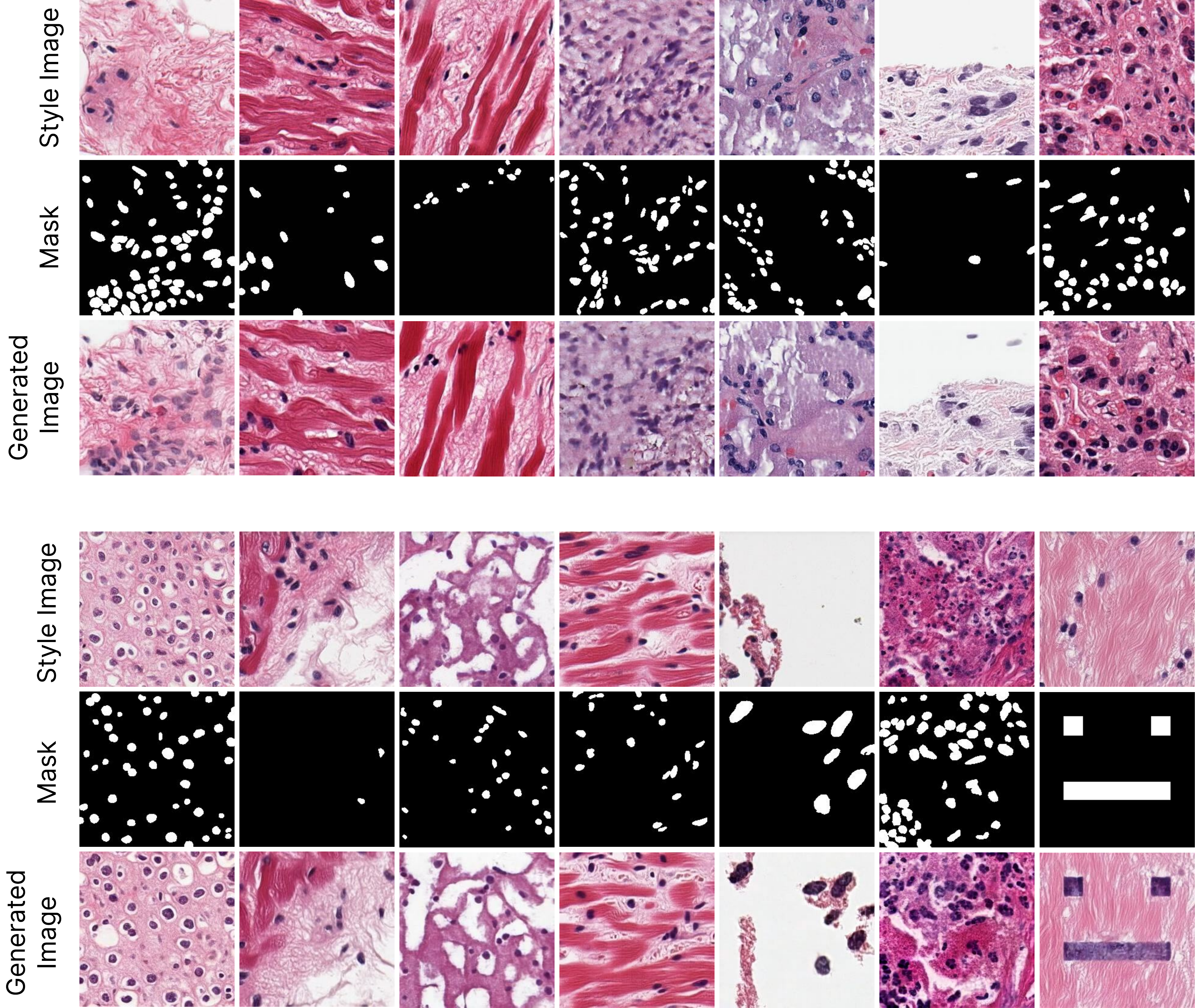}
    \caption{Images generated using the cell mask ControlNet. The synthesized samples follow the appearance of the style image, from which the UNI conditioning is extracted, and the cell layout of the reference mask. The UNI conditioning from the style image is not sterile from cell information, and when the two significantly contrast, the generated images fail to follow the mask accurately (top, columns 6-7). The guidance scale used is $w=2.5$.}
    \label{fig:controlnet_results}
\end{figure}

\begin{figure}[t]
    \centering
    \includegraphics[width=0.7\linewidth]{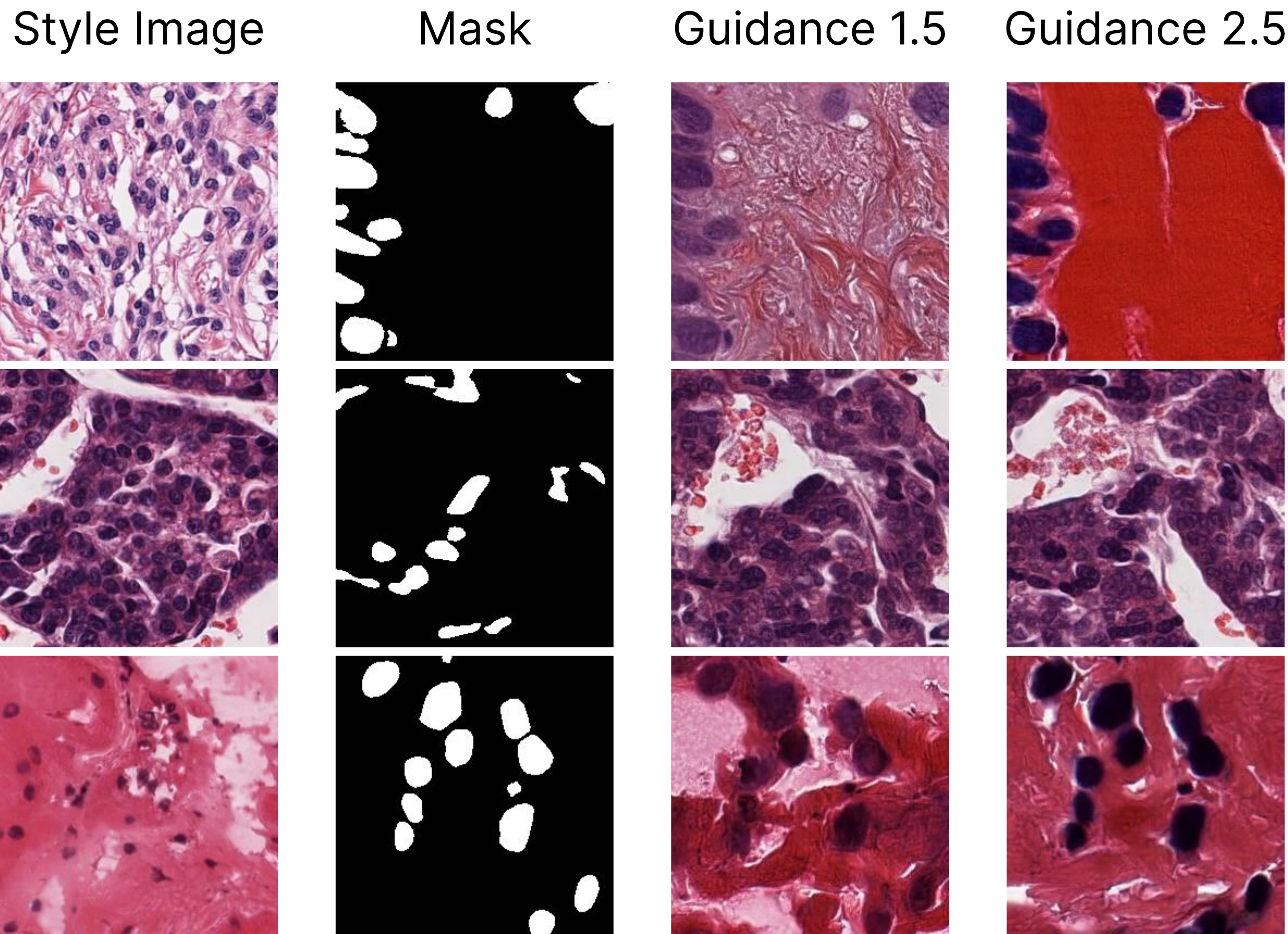}
    \caption{Effect of the guidance scale on the cell mask ControlNet generation.}
    \label{fig:controlnet_guidance}
\end{figure}

In Figure~\ref{fig:controlnet_results}, we showcase images sampled using the trained cell mask ControlNet. The PixCell-256-Cell-ControlNet model allows for finer control over the generated image. By providing the UNI embedding from a reference image and a target cell mask, we have disentangled the appearance (guided by UNI) and cell layout (guided by the mask) of the generated image. However, this disentanglement is not perfect since the UNI-2h conditioning also encodes some information about the cell count and layout. We observe that in some cases, the clashing control signals lead to images that do not exactly follow the given mask (columns 6-7 of Figure~\ref{fig:controlnet_results}). Although we can increase the guidance scale of the ControlNet conditioning, this can lead to potentially unwanted artifacts in the generated images or the model completely ignoring the UNI conditioning, as shown in Figure~\ref{fig:controlnet_guidance}.

\subsubsection{Targeted data augmentation}

Having trained the cell mask ControlNet, we set up a testbed for data augmentation using synthetic data from our model. We utilize the Lizard dataset \cite{graham2021lizard}, which provides images at $20\times$ magnification alongside their cell segmentation masks. We opt for binary cell segmentation, which our ControlNet is guided by.

As a baseline, we train a segmentation head using features from the last layer of UNI-2h. We use the convolutional segmentation head from the EVA framework \cite{kaiko.ai2024eva} that combines the extracted features with the input image to better capture cell edges. The segmentation head is trained on the Lizard data splits from GlaS, PanNuke, and CoNSeP, and is evaluated on the CRAG and DigestPath splits. We specifically selected those splits to exacerbate the difference between the training and evaluation data, testing the generalization limits of the extracted features.

We then utilize the PixCell-256-Cell-ControlNet to perform targeted data augmentation on the evaluation data. For each image in the evaluation set, we extract its UNI-2h embedding and pair it with a random training set mask. We then synthesize a new image using the two sources of conditioning. The generated image follows the appearance of the evaluation set images and has a known ground truth cell mask. We use this synthetic data as additional training samples to re-train the segmentation head. Note that this targeted data augmentation scheme does not require ground truth masks from the test set.

For the evaluation, we measure the accuracy, Dice score, and IoU between the predicted and ground truth evaluation masks. We find that with our PixCell augmentations, the accuracy increases from 0.857 \textrightarrow\  0.890, Dice score 0.629 \textrightarrow\  0.653, and IoU from 0.751 \textrightarrow\ 0.802. This improvement validates our observation that by training the cell mask ControlNet we attain finer control over the appearance and cell layout of the generated images, which we can then use to synthesize additional training data for downstream tasks.

\subsection{Additional results:}

\begin{figure}[ht]
    \centering
    \includegraphics[width=1.\linewidth]{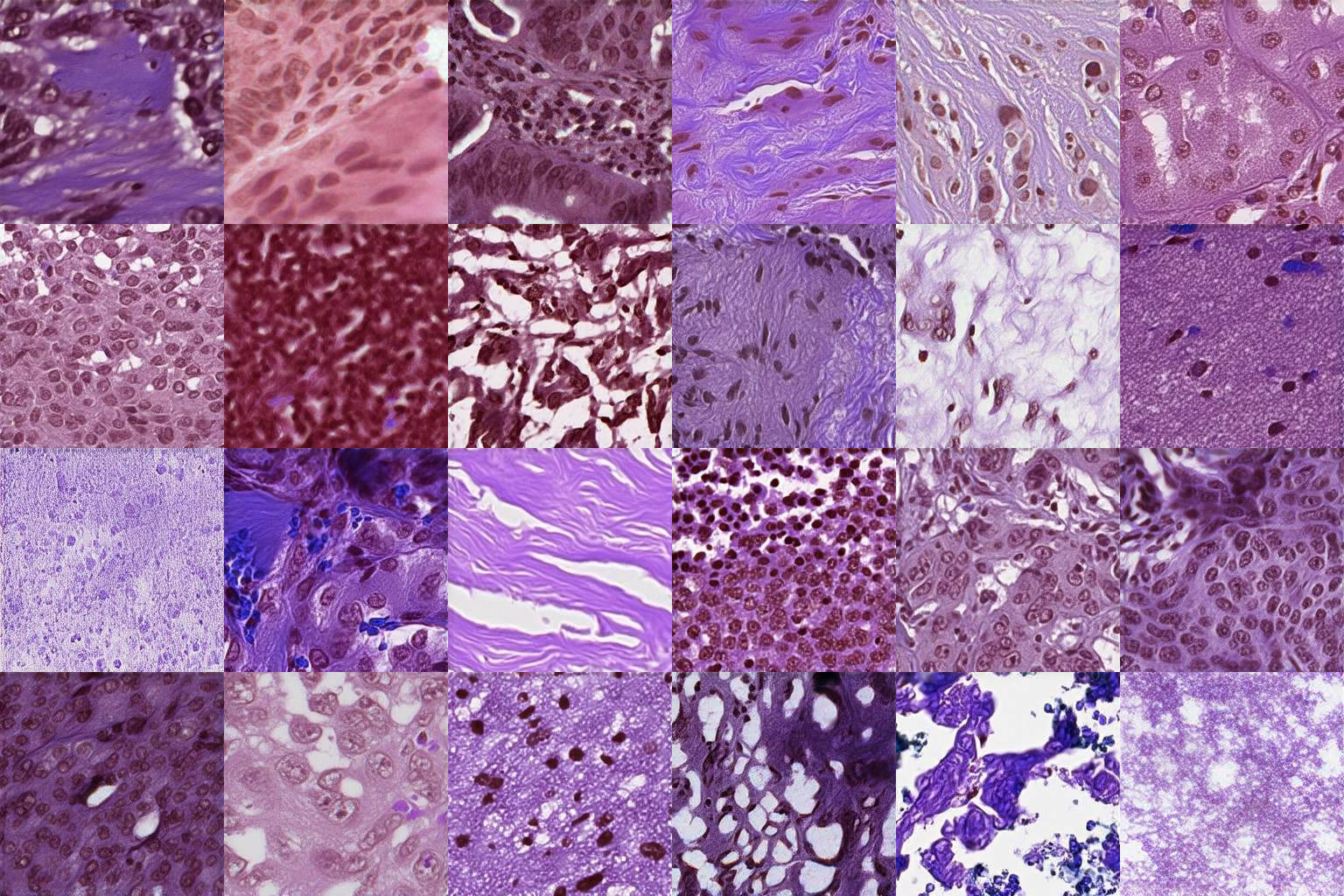}
    \caption{Samples from TCGA-Syn-10M.}
    \label{fig:synthetic_tcga}
\end{figure}

\begin{figure}[ht]
    \centering
    \includegraphics[width=1.\linewidth]{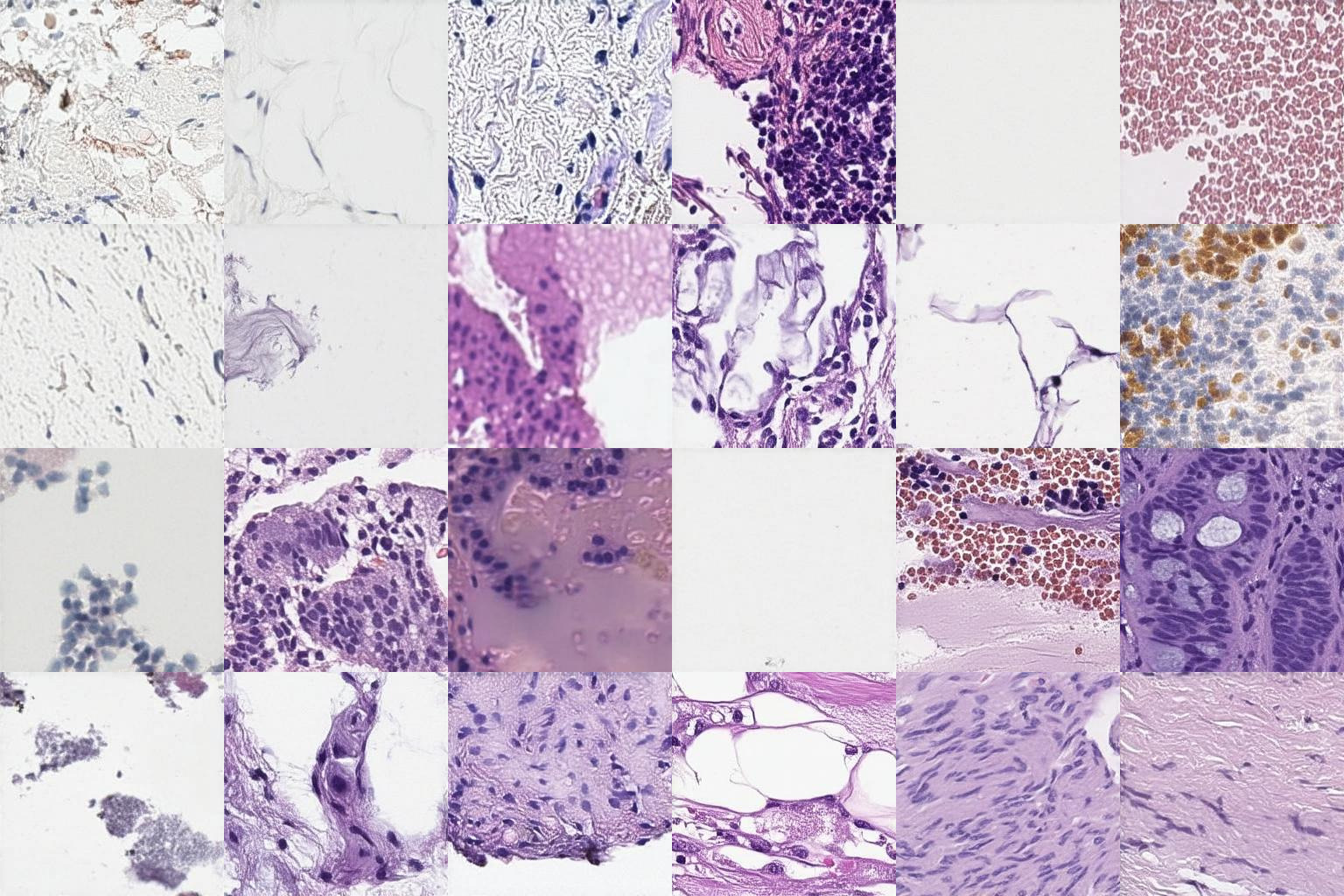}
    \caption{Samples from HistAI-Syn-10M.}
    \label{fig:synthetic_histai}
\end{figure}

\paragraph{Synthetic data} In Figure~\ref{fig:synthetic_tcga}, we show example images from TCGA-Syn-10M, the synthetic variant of TCGA generated with PixCell. Additionally, in Figure~\ref{fig:synthetic_histai} we present examples from HistAI-Syn-10M, the synthetic variant of HistAI generated with PixCell. In both cases, we find that the generated images capture a wide array of variations seen in real images, ranging from different staining types to inaccuracies in scanning (blur). By being able to generate synthetic data that closely follow the distribution of the reference set, we enable purely synthetic and synthetic-augmented self-supervised learning pipelines, which have the potential to reduce the regulatory barriers in data sharing between institutions.

\begin{figure}[ht]
    \centering
    \includegraphics[width=0.8\linewidth]{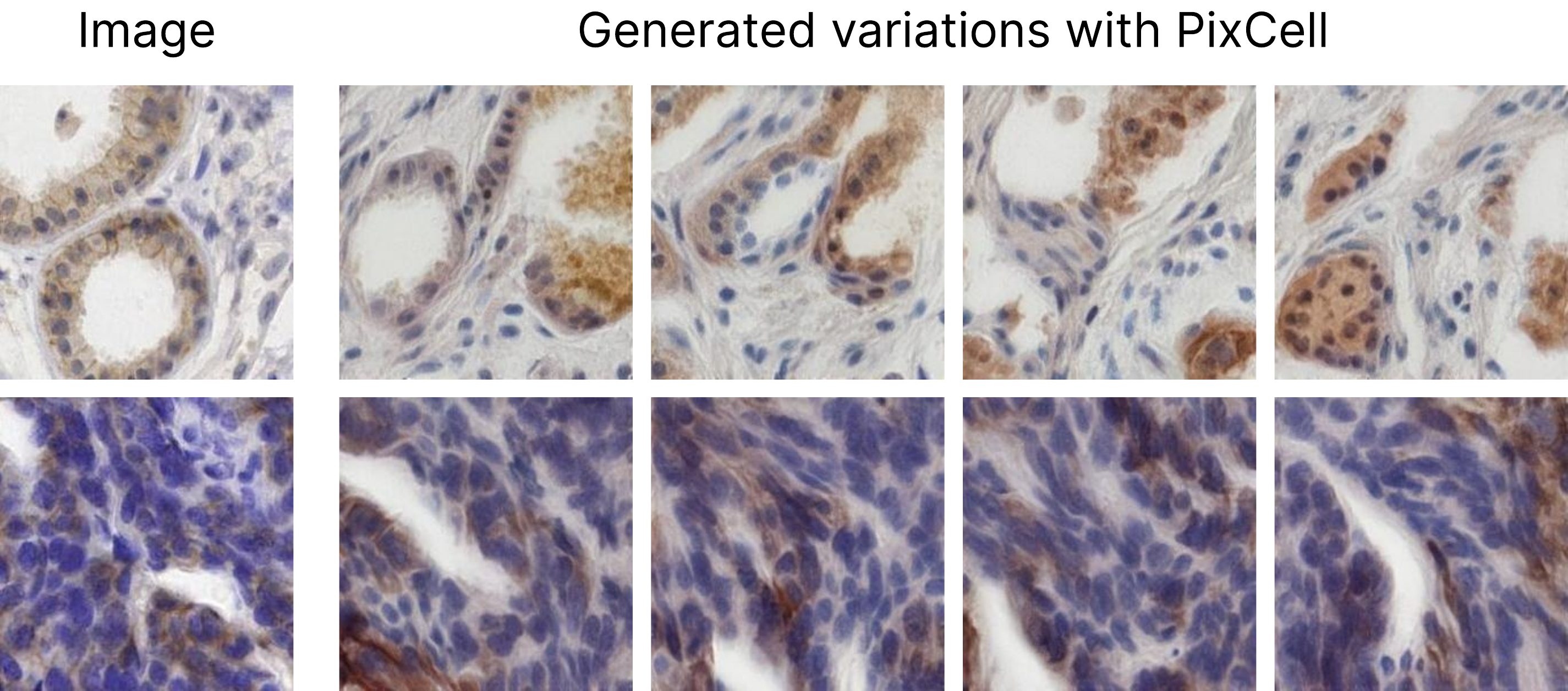}
    \caption{Using an out-of-distribution IHC image as reference, we generate variations using PixCell-256. Although never trained on IHC data, our model can synthesize images from this unseen domain.}
    \label{fig:pixcell_ihc_variations}
\end{figure}

\paragraph{Out-of-distribution generalization} In Figure~\ref{fig:pixcell_ihc_variations}, we demonstrate the generalization capabilities of PixCell-256 by generating IHC image patches that were unseen during training. By showing that PixCell can generate patches from unseen distributions, we argue that our large-scale generative pre-training has enabled the model to apply to a variety of downstream tasks, with little to no adaptation required. This supports our claim for PixCell as a generative foundation model in the digital histopathology domain.

\begin{figure}[ht]
    \centering
    \includegraphics[width=0.92\linewidth]{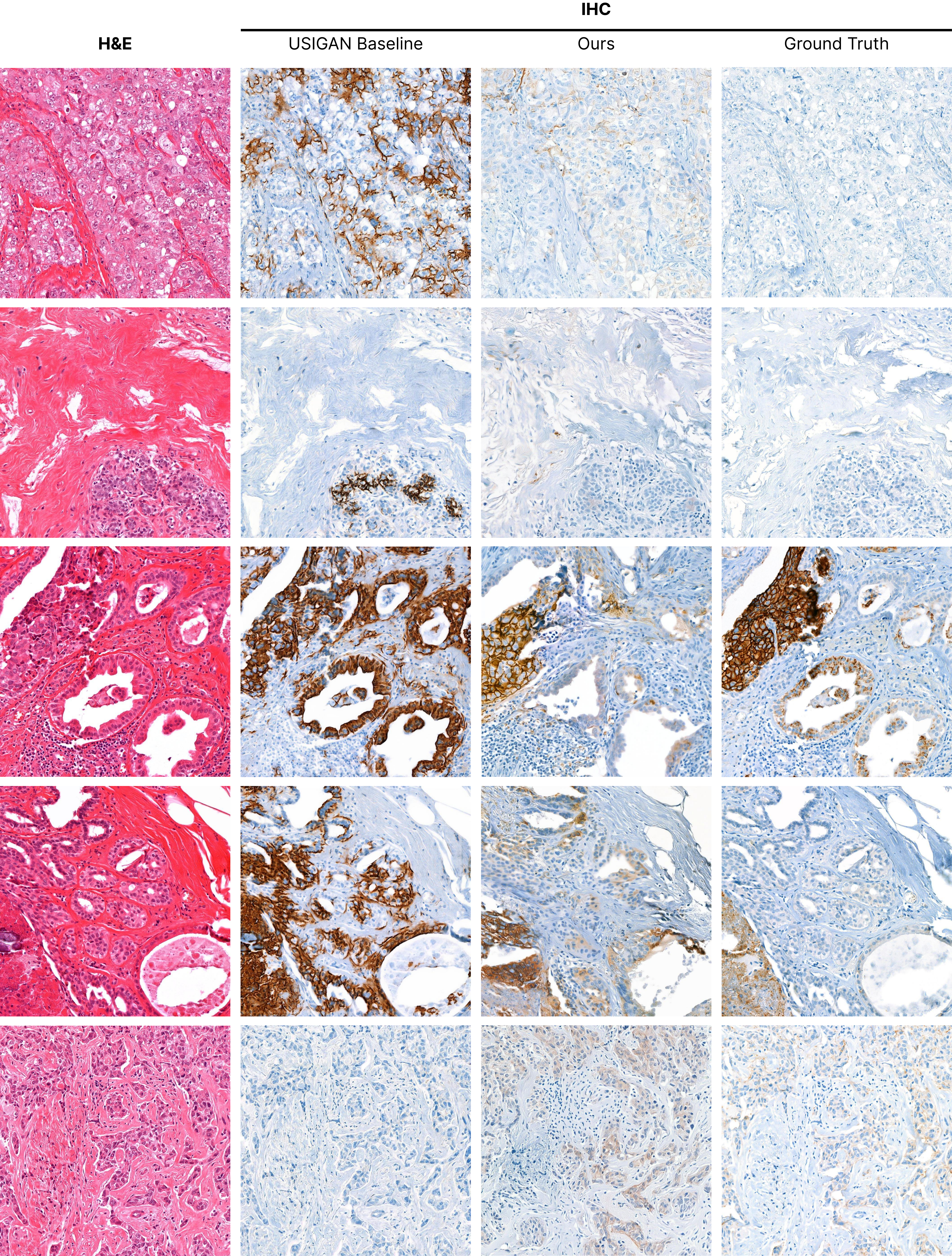}
    \caption{HER2Match virtual staining examples. We observe that the HER2Match dataset includes more negative regions than MIST, making it a more difficult dataset to apply virtual staining on. Consistent with this observation, both models have a higher false positive rate, with the baseline of USIGAN greatly exaggerating the HER2 expression in the generated images.}
    \label{fig:virtual_staining_her2match}
\end{figure}

\paragraph{Virtual staining} In Figure~\ref{fig:virtual_staining_her2match}, we present additional results of our virtual staining approach, and the corresponding result from the CycleGAN baseline USIGAN \cite{peng2025usigan}. HER2Match is a more difficult dataset than MIST, capturing a wider variety of IHC-stained tissue regions. We find that the CycleGAN baseline tends to over-stain the given tissue regions, being biased towards producing positive results. PixCell better captures the target distribution, producing results that are better-algined with the real stained tissue.

\subsection{Training hardware}
\label{sec:appendix_training_details}
We use an NVIDIA DGX A100-based cluster for our training, image quality evaluation, and self-supervised learning experiments. The DGX A100 node comprises eight NVIDIA A100 GPUs providing 320 GB of memory, two AMD EPYC 7742 CPUs with 256 cores and 1 TB of DDR4 memory. The machine uses an SSD for data storage, which offers up to 25 GB/s in bandwidth. In total, we used 3000 node hours for the paper, including exploratory experiments. 

For the controllable generation and stain translation experiments, we used NVIDIA RTX A5000 GPUs with 24GB of memory. Training ControlNets and LoRA-based adapters on our PixCell models can be done on as little as a single A5000 GPU, a compute budget trivial in comparison to the PixCell training requirements. We will be publicly releasing the ControlNet and LoRA training scripts.

\subsection{Dataset details}
\begin{figure}[ht]
    \centering
    \includegraphics[width=0.85\linewidth]{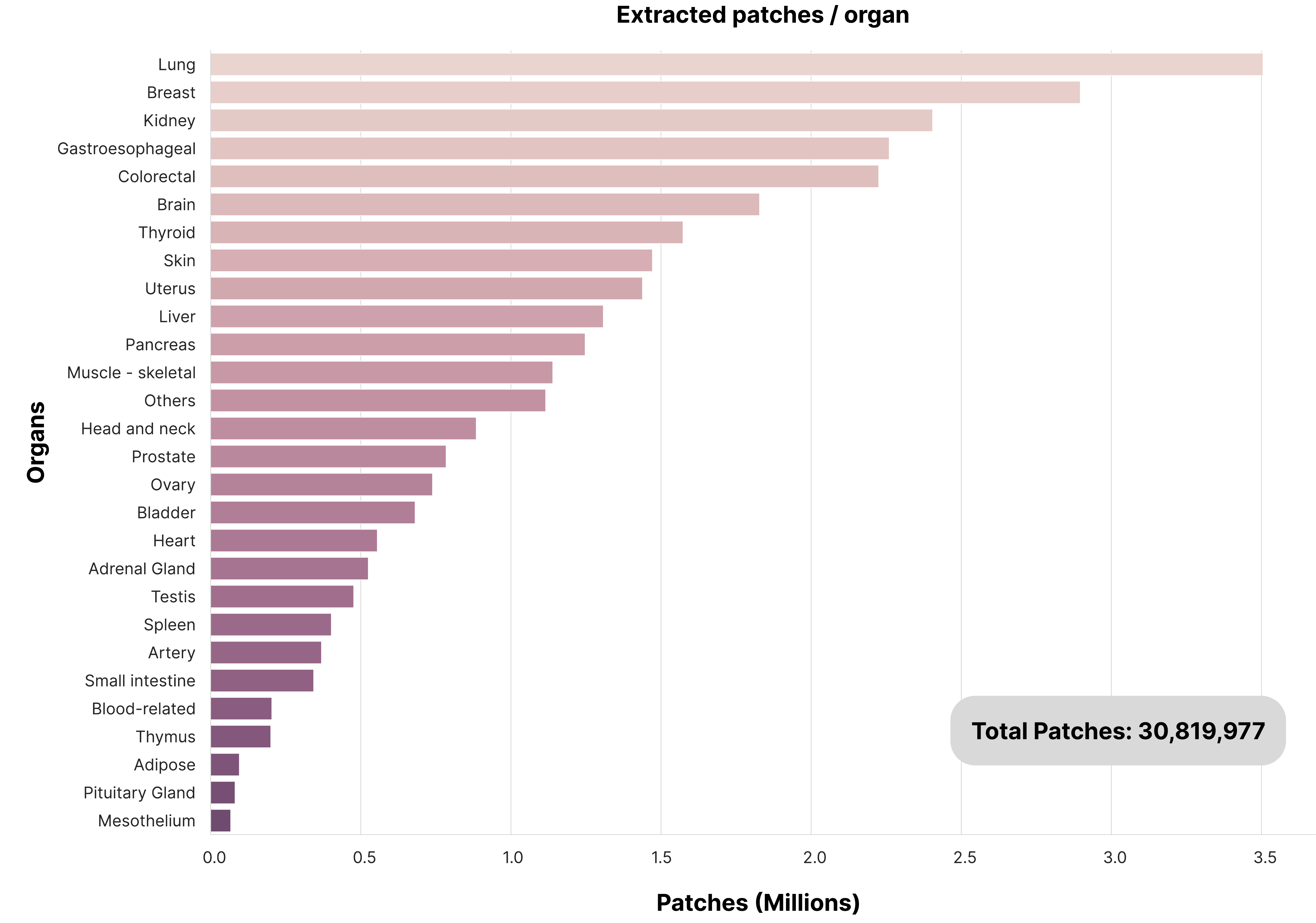}
    \caption{Distribution of organs after the $1024\times1024$ patch extraction.}
    \label{fig:data_patches}
\end{figure}

In Figure~\ref{fig:data_patches}, we visualize the distribution of organ types in the PanCan-30M dataset after patch extraction. In Table~\ref{tab:dataset_detailed}, we present the detailed composition of PanCan-30M, including organ types, data sources, dataset splits, and number of WSIs and patches per data source.

\input{tables/data_sources_full}

%% file: tables/fid_1024.tex
\begin{tabular}{cccc}
    \toprule
    Method & Time / img (s) & Crop FID \textdownarrow & CLIP FID \textdownarrow \\
    \midrule
    LRDM \cite{cvpr24_ldm} & 60 & 15.51 & 7.43 \\
    $\infty$-Brush \cite{inftybrush} & 30 & 17.87 & 3.74 \\
    ZoomLDM \cite{zoomldm} & 28 & 14.94 & 1.23 \\
    PixCell-1024 & \textbf{2.5} & \textbf{7.92} & \textbf{0.68} \\
    \bottomrule
\end{tabular}

%% file: tables/cfg_scale.tex
\begin{tabular}{ccccccccc}
    \toprule
    $w$ & 1 & 1.2 & 1.4 & 1.6 & 1.8 & 2 & 3 & 5 \\
    \midrule
    PixCell-256 & 13.25 & 12.21 & 11.16 & 10.67 & 10.34 & \textbf{9.65} &  9.81 & 10.54 \\
    PixCell-1024 & 8.39 & \textbf{7.92} & 8.55 & 9.14 & 9.72 & 10.83 & 15.2 & 20.59 \\
    \bottomrule
\end{tabular}

%% file: tables/data_augmentation_knn.tex
\begin{tabular}{ccccccc}
\toprule
               & \multicolumn{3}{c}{BACH}        & \multicolumn{3}{c}{BRACS}         \\
               \cmidrule(lr){2-4}                  \cmidrule(lr){5-7}
               & UNI-2h & Virchow-2 & Hoptimus-1 &  UNI-2h & Virchow-2 & Hoptimus-1  \\
\midrule
k-NN           & 79.4  & 74.5      & 59.7        & 52.3  & 51.0        & 51.1  \\
Linear probing & 87.0  & 82.8      & 75.1        & 63.9  & 61.7        & 64.7  \\ 
Augmentation   & 87.9  & 86.4      & 79.6        & 65.1  & 64.5        & 63.0  \\
\\
               & \multicolumn{3}{c}{Break-his}   & \multicolumn{3}{c}{mHist}        \\
               \cmidrule(lr){2-4}                  \cmidrule(lr){5-7}
               & UNI-2h & Virchow-2 & Hoptimus-1 & UNI-2h & Virchow-2 & Hoptimus-1  \\
\midrule
k-NN           & 74.1  & 75.4       & 75.0       & 67.8  & 70.7       & 71.9       \\
Linear probing & 78.0  & 76.2       & 75.6       & 80.4  & 84.2       & 81.7       \\
Augmentation   & 86.8  & 81.9       & 83.3       & 81.8  & 83.9       & 82.4       \\
\bottomrule
\end{tabular}

%% file: tables/knn_ssl_supp.tex
\resizebox{\textwidth}{!}{
\begin{tabular}{c|c|c|c|cccc|c|c|c}
\toprule
\multirow{2}{*}{Model} & \multirow{2}{*} {\begin{tabular}[c]{@{}c@{}}bracs \\ ~\cite{bracs}\end{tabular}}  & \multirow{2}{*} {\begin{tabular}[c]{@{}c@{}}ccrcc \\ ~\cite{brummer2023computational}\end{tabular}} & \multirow{2}{*}{\begin{tabular}[c]{@{}c@{}}crc \\ ~\cite{nct-crc}\end{tabular}} & \multicolumn{4}{c|}{SPIDER ~\cite{nechaev2025spider}} & \multirow{2}{*}{\begin{tabular}[c]{@{}c@{}}tcga crc \\ msi ~\cite{kather2020histological}\end{tabular}} & \multirow{2}{*}{\begin{tabular}[c]{@{}c@{}}tcga \\ tils ~\cite{komura2020histology}\end{tabular}} & \multirow{2}{*}{Avg} \\
                            &                        &                        &                      & Breast & Colorec & Skin  & Thorax &                                                                          &                            &                      \\ \midrule
TCGA Real 100\%             & 53.03                  & 80.48                  & 91.54                & 77.19  & 82.43   & 85.13 & 86.81  & 59.79                                                                    & 84.59                      & 77.89                \\
TCGA Real 100\% + Syn 100\% & 50.92                  & 75.56                  & 91.73                & 76.61  & 82.20   & 84.91 & 87.32  & 58.21                                                                    & 83.36                      & 76.76                \\ 
TCGA Real 50\%              & 50.30                  & 70.07                  & 90.75                & 77.26  & 83.79   & 84.92 & 88.08  & 58.48                                                                    & 82.33                      & 76.22                \\
TCGA Real 50\% + Syn 50\%   & 50.66                  & 74.78                  & 91.09                & 76.37  & 83.51   & 84.78 & 89.24  & 58.65                                                                    & 82.79                      & 76.87                \\\bottomrule
\end{tabular}
}

%% file: tables/data_sources_full.tex
\begin{table}[ht]
\centering
\label{tab:dataset_detailed}
\caption{Detailed description of the pretraining dataset used in training the PixCell-256 and PixCell-1024 models.}

\begin{tabular}{ccccc}
\toprule
Organ & Dataset & Split & No of WSIs & No of patches \\ \midrule
\multirow{2}{*}{Adipose} & GTEx & Adipose - Subcutaneous & 978 & 57,791 \\
 & GTEx & Adipose - Visceral (Omentum) & 815 & 37,763 \\ \midrule
\multirow{3}{*}{Adrenal Gland} & TCGA FFPE & acc & 227 & 250,571 \\
 & TCGA FF & acc & 96 & 11,322 \\
 & GTEx & Adrenal Gland & 717 & 263,138 \\ \midrule
\multirow{3}{*}{Artery} & GTEx & Artery - Aorta & 858 & 274,691 \\
 & GTEx & Artery - Coronary & 662 & 33,361 \\
 & GTEx & Artery - Tibial & 979 & 61,050 \\ \midrule
\multirow{4}{*}{Bladder} & GTEx & Bladder & 130 & 61,382 \\
 & TCGA FFPE & blca & 457 & 459,281 \\
 & TCGA FF & blca & 469 & 79,797 \\
 & SBU & Bladder & 57 & 80,421 \\ \midrule
\multirow{3}{*}{Blood-related} & TCGA FFPE & dlbc & 44 & 26,340 \\
 & TCGA FF & dlbc & 59 & 12,427 \\
 & CPTAC & AML & 93 & 164,842 \\ \midrule
\multirow{8}{*}{Brain} & GTEx & Brain - Cerebellum & 426 & 154,674 \\
 & GTEx & Brain - Cortex & 428 & 149,441 \\
 & CPTAC & GBM & 462 & 99,722 \\
 & TCGA FFPE & lgg & 844 & 563,253 \\
 & TCGA FF & lgg & 728 & 125,824 \\
 & TCGA FFPE & gbm & 860 & 492,950 \\
 & TCGA FF & gbm & 1193 & 228,187 \\
 & SBU & Brain & 15 & 14,253 \\ \midrule
\multirow{5}{*}{Breast} & GTEx & Breast - Mammary Tissue & 894 & 144,924 \\
 & CPTAC & brca & 653 & 162,981 \\
 & TCGA FFPE & brca & 1133 & 735,448 \\
 & TCGA FF & brca & 1978 & 312,557 \\
 & SBU & Breast & 1895 & 1,540,837 \\ \midrule
\multirow{11}{*}{Colorectal} & TCGA FFPE & coad & 459 & 267,397 \\
 & TCGA FF & coad & 983 & 196,568 \\
 & TCGA FFPE & read & 165 & 76,709 \\
 & TCGA FF & read & 365 & 72,122 \\
 & GTEx & Colon - Sigmoid & 800 & 281,009 \\
 & GTEx & Colon - Transverse & 937 & 303,477 \\
 & CPTAC & coad & 372 & 126,416 \\
 & SBU & Appendix & 199 & 149,278 \\
 & SBU & Cecum & 61 & 5,544 \\
 & SBU & Colon & 354 & 546,203 \\
 & SBU & Rectum & 295 & 200,305 \\ \midrule
\multirow{11}{*}{Gastroesophageal} & TCGA FFPE & esca & 158 & 114,285 \\
 & TCGA FF & esca & 238 & 35,297 \\
 & TCGA FFPE & stad & 442 & 305,349 \\
 & TCGA FF & stad & 755 & 271,865 \\
 & GTEx & Stomach & 939 & 391,985 \\
 & GTEx & Esophagus - Gastroesophageal & 787 & 363,692 \\
 & GTEx & Esophagus - Mucosa & 963 & 211,581 \\
 & GTEx & Esophagus - Muscularis & 958 & 420,428 \\
 & SBU & Esophagus & 101 & 19,980 \\
 & SBU & Gastroesophageal Junction & 2 & 3 \\
 & SBU & Stomach & 154 & 125,812 \\ \bottomrule
\end{tabular}
\end{table}

\begin{table}[ht]
\centering
\begin{tabular}{ccccc}
\toprule
Organ & Dataset & Split & No of WSIs & No of patches \\ \midrule
\multirow{13}{*}{Head and neck} & CPTAC & HNSCC & 390 & 112,431 \\
 & TCGA FFPE & hnsc & 472 & 312,557 \\
 & TCGA FF & hnsc & 791 & 79,785 \\
 & GTEx & Minor Salivary Gland & 247 & 23,017 \\
 & SBU & Epiglottis & 1 & 1 \\
 & SBU & Gingiva & 3 & 3,510 \\
 & SBU & Larynx & 171 & 133,126 \\
 & SBU & Lip & 44 & 49,189 \\
 & SBU & Maxilla & 63 & 46,149 \\
 & SBU & Parotid & 53 & 45,155 \\
 & SBU & Submandibular Gland & 9 & 15,571 \\
 & SBU & Tongue & 10 & 41,621 \\
 & SBU & Tonsil & 6 & 22,393 \\ \midrule
\multirow{2}{*}{Heart} & GTEx & Heart - Atrial Appendage & 614 & 246,393 \\
 & GTEx & Heart - Left Ventricle & 761 & 308,624 \\ \midrule
\multirow{10}{*}{Kidney} & GTEx & Kidney - Cortex & 557 & 223,111 \\
 & GTEx & Kidney - Medulla & 49 & 19,411 \\
 & TCGA FFPE & kich & 121 & 104,411 \\
 & TCGA FF & kich & 205 & 86,226 \\
 & TCGA FFPE & kirc & 519 & 449,337 \\
 & TCGA FF & kirc & 1654 & 494,982 \\
 & TCGA FFPE & kirp & 300 & 242,237 \\
 & TCGA FF & kirp & 473 & 103,243 \\
 & CPTAC & CCRCC & 884 & 192,805 \\
 & SBU & Kidney & 112 & 489,008 \\ \midrule
\multirow{8}{*}{Liver} & TCGA FFPE & chol & 39 & 48,090 \\
 & TCGA FF & chol & 71 & 13,171 \\
 & TCGA FFPE & lihc & 379 & 335,514 \\
 & TCGA FF & lihc & 491 & 82,474 \\
 & GTEx & Liver & 610 & 272,481 \\
 & SBU & Bile Duct & 56 & 1,823 \\
 & SBU & Gallbladder & 53 & 49,131 \\
 & SBU & Liver & 259 & 505,170 \\ \midrule
\multirow{9}{*}{Lung} & TCGA FFPE & luad & 541 & 398,649 \\
 & TCGA FF & luad & 1067 & 209,038 \\
 & TCGA FFPE & lusc & 512 & 380,402 \\
 & TCGA FF & lusc & 1100 & 210,390 \\
 & GTEx & Lung & 860 & 297,908 \\
 & CPTAC & LUAD & 969 & 520,881 \\
 & CPTAC & LSCC & 1018 & 488,212 \\
 & SBU & Lung & 5 & 630 \\
 & others & NLST & 1225 & 999,154 \\ \midrule
\multirow{3}{*}{Mesothelium} & TCGA FFPE & meso & 87 & 52,567 \\
 & TCGA FF & meso & 88 & 14,085 \\
 & SBU & Peritoneum & 1 & 1 \\ \midrule
\multirow{5}{*}{Muscle - skeletal} & CPTAC & SAR & 305 & 95,257 \\
 & TCGA FFPE & sarc & 600 & 639,211 \\
 & TCGA FF & sarc & 290 & 48,382 \\
 & GTEx & Muscle - Skeletal & 1001 & 356,578 \\
 & SBU & Stomach & 154 & 125,812 \\ \bottomrule
\end{tabular}
\end{table}

\begin{table}[ht]
\centering
\begin{tabular}{ccccc}
\toprule
Organ & Dataset & Split & No of WSIs & No of patches \\ \midrule
\multirow{16}{*}{Other Types} & TCGA FFPE & pcpg & 196 & 184,298 \\
 & TCGA FF & pcpg & 189 & 25,688 \\
 & GTEx & Nerve - Tibial & 975 & 95,623 \\
 & SBU & Bone & 29 & 51,297 \\
 & SBU & Cervix & 2 & 11,402 \\
 & SBU & Fallopian Tube & 7 & 320 \\
 & SBU & Lymph Node & 238 & 442,527 \\
 & SBU & Mediastinum & 18 & 22,158 \\
 & SBU & Mesentery & 9 & 1 \\
 & SBU & Negative & 11 & 2,303 \\
 & SBU & Omentum & 10 & 148 \\
 & SBU & Peripheral Nerve & 6 & 1,076 \\
 & SBU & Soft Tissue & 61 & 143,760 \\
 & SBU & Spinal Cord & 5 & 1,387 \\
 & SBU & Unknown & 102 & 106,210 \\
 & SBU & Ureter & 17 & 27,609 \\ \midrule
\multirow{5}{*}{Ovary} & TCGA FFPE & ov & 107 & 102,005 \\
 & TCGA FF & ov & 1374 & 372,215 \\
 & GTEx & Ovary & 255 & 105,532 \\
 & SBU & Ovary & 17 & 15,135 \\
 & others & PTRC-HGSOC & 348 & 144,134 \\ \midrule
\multirow{6}{*}{Pancreas} & TCGA FFPE & paad & 209 & 168,177 \\
 & TCGA FF & paad & 257 & 42,614 \\
 & GTEx & Pancreas & 865 & 374,094 \\
 & CPTAC & PDA & 567 & 98,216 \\
 & SBU & Ampulla & 142 & 90,211 \\
 & SBU & Pancreas & 449 & 473,837 \\ \midrule
Pituitary Gland & GTEx & Pituitary & 428 & 81,116 \\ \midrule
\multirow{4}{*}{Prostate} & GTEx & Prostate & 599 & 262,357 \\
 & TCGA FFPE & prad & 449 & 363,268 \\
 & TCGA FF & prad & 723 & 139,815 \\
 & SBU & Prostate & 5 & 18,765 \\ \midrule
\multirow{8}{*}{Skin} & TCGA FFPE & uvm & 80 & 46,105 \\
 & TCGA FF & uvm & 70 & 8,454 \\
 & TCGA FFPE & skcm & 475 & 387,694 \\
 & TCGA FF & skcm & 475 & 75,806 \\
 & GTEx & Skin - Not Sun Exposed & 818 & 333,848 \\
 & GTEx & Skin - Sun Exposed & 1001 & 359,139 \\
 & CPTAC & CM & 411 & 138,346 \\
 & SBU & Skin & 116 & 122,049 \\ \midrule
\multirow{4}{*}{Small intestine} & GTEx & Small Intestine - Terminal & 798 & 180,567 \\
 & SBU & Duodenum & 56 & 22,758 \\
 & SBU & ileum & 12 & 9,905 \\
 & SBU & Small Bowel & 99 & 129,820 \\ \midrule
\multirow{2}{*}{Spleen} & GTEx & Spleen & 874 & 376,920 \\
 & SBU & Spleen & 5 & 24,999 \\ \midrule
\multirow{4}{*}{Testis} & TCGA FFPE & tgct & 254 & 240,765 \\
 & TCGA FF & tgct & 159 & 27,182 \\
 & GTEx & Testis & 592 & 208,431 \\
 & SBU & Stomach & 154 & 125,812 \\ \bottomrule
\end{tabular}
\end{table}

\begin{table}[ht]
\centering
\begin{tabular}{ccccc}
\toprule
Organ & Dataset & Split & No of WSIs & No of patches \\ \midrule
\multirow{4}{*}{Thyroid} & TCGA FFPE & thca & 519 & 398,380 \\
 & TCGA FF & thca & 639 & 94,571 \\
 & GTEx & Thyroid & 902 & 283,835 \\
 & SBU & Thyroid & 285 & 796,642 \\ \midrule
\multirow{13}{*}{Uterus} & GTEx & Cervix - Ectocervix & 38 & 13,062 \\
 & GTEx & Cervix - Endocervix & 42 & 14,578 \\
 & GTEx & Fallopian Tube & 43 & 4,979 \\
 & GTEx & Uterus & 237 & 106,294 \\
 & GTEx & Vagina & 276 & 137,860 \\
 & TCGA FFPE & ucs & 91 & 99,858 \\
 & TCGA FF & ucs & 63 & 9,524 \\
 & TCGA FFPE & cesc & 279 & 175,492 \\
 & TCGA FF & cesc & 325 & 49,798 \\
 & TCGA FFPE & ucec & 566 & 601,015 \\
 & TCGA FF & ucec & 805 & 143,477 \\
 & SBU & Uterus & 16 & 82,294 \\
 & SBU & Vagina & 1 & 1 \\ \bottomrule 
\end{tabular}
\end{table}